\documentclass[prd,preprintnumbers,amssymb,amsmath,12pt,nofootinbib,longbibliography]{revtex4}
\usepackage{graphicx,hyperref,multirow,subfigure}
\usepackage{color}

\begin{document}

\title{Interference effects in heavy $W^\prime$-boson searches at the LHC}

\date{\today}

\author{Elena Accomando}%
 \email{e.accomando@soton.ac.uk}
\affiliation{NExT Institute and School of Physics and Astronomy, University of Southampton, Highfield,
Southampton SO17 1BJ, UK}%
\author{Diego Becciolini}%
 \email{diego.becciolini@soton.ac.uk}
 \affiliation{NExT Institute and School of Physics and Astronomy, University of Southampton, Highfield,
Southampton SO17 1BJ, UK}%
\author{Stefania De Curtis}%
 \email{decurtis@fi.infn.it}
\affiliation{Istituto Nazionale di Fisica Nucleare, Sezione di Firenze - Italy}%
\author{Daniele Dominici}%
 \email{dominici@fi.infn.it}
\affiliation{Universit\`a degli Studi di Firenze, Dipartimento di
Fisica e Astronomia, Firenze - Italy
and Istituto Nazionale di Fisica Nucleare, Sezione di Firenze - Italy}%
 \author{Luca Fedeli}%
 \email{fedeli@fi.infn.it}
\affiliation{Universit\`a degli Studi di Firenze, Dipartimento di
Fisica e Astronomia, Firenze - Italy
and Istituto Nazionale di Fisica Nucleare, Sezione di Firenze - Italy}%
\author{Claire Shepherd-Themistocleous}%
 \email{claire.shepherd@stfc.ac.uk}
\affiliation{Particle Physics Department,
STFC, Rutherford Appleton Laboratory, Harwell Science and Innovation Campus,
Didcot, Oxfordshire, OX11 0QX, UK}%

\begin{abstract}
Interference effects are widely neglected in searches for new physics.
This is the case in recent publications on searches for $W^\prime$-bosons using
leptonic final states.

We examine the effects of interference on distributions frequently used to
determine mass limits for possible $W^\prime$-bosons and show that there are
important \emph{qualitative} effects on the behaviour of the new physics signal.

There are two main consequences.
Firstly, exclusion limits where interferences effects have not been considered are likely to have been overestimated.
Secondly, presenting experimental results as a function of a transverse mass cut rather than in terms of the contribution of new physics to the total cross-section would be more informative.
\end{abstract}

\maketitle

\section{Introduction}

Heavy charged $W^\prime$-bosons arise in a number of theories that extend the Standard Model (SM) gauge group
\footnote{$W^\prime$-bosons arise less frequently than $Z^\prime$-bosons as the extension needs to contain a $SU(2)$ factor, while for the latter a simple $U(1)$ factor is sufficient.
For a recent review on $Z^\prime$ scenarios, see \cite{Accomando:2010fz}}.
The Left-Right symmetric (LR) class of models
\cite{Pati:1974yy,Mohapatra:1975jm,Senjanovic:1975rk},
based on the enlarged symmetry $SU(2)_L\otimes SU(2)_R\otimes U(1)$, is an old and popular example;
within these models the discovery reach of the LHC has been recently re-investigated
\cite{Gopalakrishna:2010xm,Frank:2010cj,Maiezza:2010ic}
and bounds have been derived using results published by the LHC collaborations \cite{Nemevsek:2011hz}.
A second class is represented by extra-dimensional theories
\cite{ArkaniHamed:1998rs,Antoniadis:1998ig,Randall:1999ee},
where $W^\prime$-bosons emerge as Kaluza-Klein excitations of the SM gauge bosons.
Within the ADD model, the phenomenology of signals from extra charged gauge bosons at the LHC have been discussed in Refs. \cite{Accomando:1999sj,Boos:2011ib}.
In the RS1 model with gauge bosons in the 5D bulk and fermions on the UV brane, analogous results are published in Refs.
\cite{Agashe:2007ki,Agashe:2008jb}.
Five-dimensional models can also be deconstructed to the usual four-dimensional space-time
\cite{ArkaniHamed:2001ca,ArkaniHamed:2001nc,Hill:2000mu,Cheng:2001vd,Abe:2002rj,Falkowski:2002cm,Randall:2002qr,Son:2003et,deBlas:2006fz},
where they are described by chiral Lagrangians with extended gauge symmetries.
Within this framework, Higgsless theories find their natural 5D interpretation.
The simplest deconstructed model is called minimal Higgsless model (or 3-site in the linear moose language),
and represents the 5D interpretation of the old BESS model \cite{Casalbuoni:1985kq,Casalbuoni:1986vq}.
The tension between ElectroWeak Precision Test (EWPT) and unitarity requirements constrain the ${W^\prime}^\pm$-boson to be almost fermiophobic
\cite{Casalbuoni:2005rs,Chivukula:2005ji,Chivukula:2005cc,Bechi:2006sj,Casalbuoni:2007xn}.
Its next-to-minimal extension (or 4-site model) described in Refs.
\cite{Accomando:2008jh,Accomando:2008dm,Accomando:2010fz,Accomando:2010ir,Accomando:2011vt,Accomando:2011xi}
relaxes the above-mentioned dichotomy thus allowing sizeable couplings between extra ${W^\prime}^\pm$-bosons and SM fermions as predicted by more general extra-dimensional theories.
Technicolor models represent the last class of theories implying the existence of new heavy charged gauge bosons \cite{Andersen:2011yj}.
This class, which historically provides an alternative electroweak symmetry breaking (EWSB) mechanism to the Higgs procedure, predicts multiple extra ${W^\prime}^\pm$ (and $Z^\prime$) bosons \cite{Belyaev:2008yj}. For a recent phenomenological study, see \cite{Andersen:2011nk}.

Experimental searches for a $W^\prime$-boson at the Tevatron and the LHC are usually interpreted in the context of a benchmark scenario inspired by Ref. \cite{Altarelli:1989ff}.
This model just includes one extra charged vector boson with couplings to fermions identical to those of the corresponding SM $W$-boson, and no mixing with the EW SM bosons.
In this paper we consider two representative models: the above model, which we refer to as the ``benchmark model'', and the 4-site model
\cite{Accomando:2008jh,Accomando:2008dm,Accomando:2010ir,Accomando:2011vt,Accomando:2011xi},
which we feel is relevant in this context as fermion couplings larger than in the more minimal 3-site model can be expected.
Furthermore the phenomenology of the latter model is close to that of Technicolour-type scenarios, so it should be representative of this class of theories as well.

We use this framework to quantitatively discuss the topic of this paper, which is the magnitude of the interference terms between BSM contributions and SM background, and the importance of such terms in new physics signals.
We note that at least some experimental analyses performed by the LHC collaborations
--- the latest ones were published this summer \cite{CMS-PAS-EXO-11-024,Aad:2011yg} ---
do not include interference in computing the expected $W^\prime$-boson signal.
For these analyses Monte Carlo events are generated with PYTHIA, where the $W^\prime$ and the SM background are simulated separately.
These two contributions are then summed to represent the prediction of the model, i.e.\ their mutual interference is assumed to be negligible.
Within a generic model, this assumption is valid close to the $W^\prime$ resonance peak, but may not be elsewhere.

In this paper, we show the effect of the interference terms on the physical observables used to search for a $W^\prime$-boson.
The published results from the LHC collaborations are mainly based on analysing final states with one isolated charged lepton and a neutrino.
We will show the impact of the interference terms on the dilepton transverse mass distribution.
The effect of interference is well known and has been previously discussed \cite{Boos:2006xe}, in the context of $W^\prime$ decaying to $t b$.
Subsequent analyses of this channel have considered the interference \cite{Abazov:2008vj}, whereas analyses using other final states have not.
In this paper we focus on the leptonic channel $W^\prime \rightarrow l\nu_l$.
It should however appear clearly that, in general, interference effects can be important and their relevance should be established in particular analyses.

Similar computations of interference effects to the ones we present here, for the same process, have been performed previously \cite{Rizzo:2007xs,Papaefstathiou:2009sr}.
In the first reference, the author points out that the interference effect provides the most sensitive means of determining the helicity of a potential $W^\prime$ while in the second paper the focus is on the evaluation of Next-to-Leading order (NLO) QCD effects.
As interference is still not taken into account in on-going searches, we discuss here the consequences of accounting for this contribution in experimental analyses.

We aim to discuss more specifically experimental searches and strategies used to extract and present exclusion bounds on the $W^\prime$ mass.
In the experimental analyses performed to date, the expected signal is calculated taking into account only the pure $W^\prime$-boson contribution.
Although the search is conducted in a limited higher-energy region where the importance of the interference terms is indeed reduced,
the results are then extrapolated over the whole mass-range in order to present limits on the cross-section that are independent of any kinematic cut.
However, the interference between $W^\prime$ and SM background results in a significant change in the signal prediction:
the BSM contribution to the total cross-section can even be negative.
Thus, this latter quantity is not, for a generic model, representative of what the signal is in the search-window, and expressing the observed limit in terms of it is not the most convenient choice.
We therefore argue that the way of presenting 95\% CL upper limits on the $W^\prime$-boson production cross-section could be improved,
and that leaving in a kinematic cut on the dilepton transverse mass would be more meaningful and useful to theorists.

What we want to stress is that, unlike other effects such as NLO QCD corrections, the interference contribution varies greatly over the kinematic range; it does not simply shift the prediction by an overall factor.
In the region of interest --- close to the BSM resonance peak --- the interference is negligible as expected, therefore taking it into account should not significantly shift the value of exclusion bounds on $W^\prime$-boson masses.
Outside of this range, however, it becomes important, thus is fundamental in determining an appropriate search-window.

Although the interpretations of current searches are perfectly valid assuming a purely right-handed $W^\prime$, transposing these results to a more general case is not trivial because of the interference. We also hope to convey the message that in \emph{any} search where interference may occur and not only in this particular $W^\prime$ example, its importance has to be established.

In Section \ref{sec:interference}, we discuss the interference terms and their generic effects.
We underline their potential impact on event generation and their importance to the signal of BSM physics.
In Section \ref{sec:application}, we quantify our considerations focussing on the heavy $W^\prime$-boson search at the LHC considering, as mentioned before, a benchmark scenario based on Ref. \cite{Altarelli:1989ff} and the 4-site Higgsless model \cite{Accomando:2008jh,Accomando:2008dm,Accomando:2010fz,Accomando:2010ir,Accomando:2011vt,Accomando:2011xi},
and we analyse the extraction of the 95\% CL exclusion bounds on the $W^\prime$-boson mass.
We comment, in particular, on the way the 95\% CL upper bound on the BSM physics cross-section is presented by the experimental collaborations. 

\section{The importance of interference terms}
\label{sec:interference}

\subsection{Background, signal \& interference}

In general, when considering a specific process that occurs \emph{both} in the SM and in the BSM model under investigation, there will be both SM and BSM diagrams contributing.
This happens, in particular, if one considers a model in which SM particles have heavier partners with similar interactions (e.g.\ one or more \emph{left-handed} $W^\prime$-bosons, as \emph{right-handed} ones would not actually couple to the same states, thus would not contribute to the same matrix-element as the SM $W$-boson).
Since the probability for the process to occur depends on the total amplitude squared, there will be cross-terms between SM and BSM diagrams on top of the purely SM or BSM contributions.
In other words, the matrix element squared entering the computation of a cross-section is
\begin{equation}
|\mathcal{M}|^2 = |\mathcal{M}_{SM} + \mathcal{M}_{BSM}|^2 = |\mathcal{M}_{SM}|^2 + |\mathcal{M}_{BSM}|^2 + 2\, \mathrm{Re}\, (\mathcal{M}_{SM}^*\cdot \mathcal{M}_{BSM}).
\end{equation}
The first of these terms is the SM prediction, which constitutes the \emph{irreducible background} in the search for new physics.
The other two thus form together the \emph{signal of new physics}, since it is defined as the difference between the prediction of the hypothesised model and the SM.
The part of the signal that mixes together SM and BSM contributions is usually referred to as the \emph{interference term}.

It should be noted that, whenever one amplitude is significantly larger than the other, the corresponding squared term will dominate the whole expression; on the other hand, if both amplitudes are of the same order, all three terms of the expression will be comparable in magnitude.

To illustrate the effects of interference, we focus on a simple case of particular interest: a process in which the $s$-channel exchange of a given particle and of the corresponding heavier resonance are the only contributing diagrams.
This is what happens in the Drell-Yan production and decay of an extra $W^\prime$-boson. 
The cross-section goes like
\begin{align}
\frac{\sigma}{s} \propto & \left( \frac{a_{SM}^2}{s-m_{SM}^2} + \frac{a_{BSM}^2}{s-m_{BSM}^2} \right)^2 \nonumber\\
&=
\left( \frac{a_{SM}^2}{s-m_{SM}^2} \right)^2 +
\left( \frac{a_{BSM}^2}{s-m_{BSM}^2} \right)^2 +
2 \left( \frac{a_{SM}^2}{s-m_{SM}^2} \cdot \frac{a_{BSM}^2}{s-m_{BSM}^2} \right)
\label{eq:propag}
\end{align}
where $m_i$ are the masses of the resonances, $a_i$ the corresponding couplings to SM fermions, and $s$ is the center-of-mass energy squared.

The above-mentioned interference term is often neglected, and thus only the contribution of the extra heavy resonance alone is usually taken into account.
For instance, the current implementation of the benchmark model taken from Ref. \cite{Altarelli:1989ff} in PYTHIA does not include interference.
We will show that, although this approximation is reliable around the resonance peak, it is not valid outside the peaking region.
The validity domain of such an approximation should be checked before performing any computation and not just in the context of the charged Drell-Yan process with a $W^\prime$-boson.

\subsection{Relative sizes \& interplay}

In energy-ranges close to either the SM or the BSM resonance, provided the BSM coupling is not too small compared to the SM one (generally true for cases of interest), the background or the pure BSM term, respectively, will dominate.
However, there will be a region somewhere between the two resonances ($m_{SM}^2 < s < m_{BSM}^2$) where these two terms are of the same order
(in the zero-width limit and if $a_{SM}=a_{BSM}$, they are equal for $s=\frac{1}{2} (m_{SM}^2 + m_{BMS}^2)$).
If the pure BSM term and the background are of the same order, then the magnitude of the interference must be comparable to the other terms (in the zero-width limit, when background and BSM part are equal, the interference is twice as large as each of the other terms).

To summarise, it should be obvious from eq.\ \eqref{eq:propag} --- if the couplings are not too different from each other and the widths not too large ---
that as one approaches the SM peak, the absolute magnitudes of the three terms are ordered as follows:
\begin{equation}
 \text{SM} > \text{interf.} > \text{BSM};
\end{equation}
close to the BSM resonance, the hierarchy is inverted:
\begin{equation}
 \text{BSM} > \text{interf.} > \text{SM};
\end{equation}
but then there is an intermediate region where
\begin{equation}
 \text{interf.} > \text{BSM} \sim \text{SM}.
\end{equation}
The interference is therefore never the smallest contribution
\footnote{At both resonance peaks the interference changes sign and therefore is the smallest of all three contributions in the narrow region where this happens, whose size (in $\sqrt{s}$) is approximately ${m_i\, \Gamma_i^2}/{(2\, (m_{BSM}^2-m_{SM}^2))}$, with $m_i$ and $\Gamma_i$ being the mass and width of the corresponding resonance.}.
If one wants to consider the intermediate energy range in any way (for instance fitting to the shape of the distribution in this region), the interference cannot be neglected.
When subtracting the background from the prediction, one cannot then consider the lower-energy region without including the interference, as it dominates the signal there.

When the BSM coupling is small and/or the width of the extra resonance large, so that even at the BSM resonance peak the background is important, the effect of the interference is particularly striking:
because of the sign change at the resonance, it makes the peak more visible than one would expect when considering only the pure BSM contribution.

The other important feature of the interference one should note is that it is \emph{negative} in the entire range $m_{SM}^2 < s < m_{BSM}^2$,
unless the coupling factor appearing in the interference term is negative, which happens only if the interaction structure of the BSM particle differs from the corresponding SM one in a particular way.
In our case of interest, the $W^\prime$ would have to couple non-universally to fermions, with a sign-difference between the lepton and quark couplings.
The Un-unified Standard Model is an example of one such case \cite{Georgi:1989xz}.

We chose to focus our attention on the case where the interference is destructive between the resonances:
firstly, because it corresponds to the standard benchmark scenario;
secondly, because it leads to an important change in the behaviour of the BSM prediction, which becomes negative in a large region of interest.

When considering hadron collisions, the partonic cross-section has to be convoluted with the parton distribution functions (PDF).
Moreover, when searching for a $W^\prime$-boson in a charged lepton plus neutrino final state, the kinematic observable used is the dilepton transverse mass $m_T$, not $\sqrt{s}$.
These details do not modify the general features we have just discussed, as should be evident from the results presented in the following sections.

If more particles contribute and/or the interaction structures of the particles are not the same, then extra complications occur.
For instance, in the corresponding $Z^\prime$ process there are two SM particles contributing: the $Z$ and the photon;
these do not couple to fermions of one helicity only unlike the SM $W$, and the contributions from the down-type and up-type quark processes are in general different.
As a result, in the large variety of $Z^\prime$ models \cite{Accomando:2010fz}, one observes differences in the relative size and sign of the interference.
In the present paper we focus on the charged Drell-Yan channel with SM-like $W^\prime$-bosons (i.e.\ destructive interference), and will confine any discussion of the neutral case to future work.

\subsection{Problem with event generation}
\label{sec:generation}

The inclusion of the interference term in computing distributions and cross-sections is easy from the theoretical point of view,
especially for 2 $\rightarrow$ 2 particle processes like the Drell-Yan production and decay of $W^\prime$-bosons.
The problem arises only when generating the expected number of events in a complete simulation.

When investigating different BSM scenarios in detail, and generating events for each case, one would wish to separately consider the SM part, which is common to all, and the ``model dependent'' BSM prediction.
If the two contributions to the cross-section are both positive, it is straightforward to independently produce background and new physics events, and then add these up using appropriate re-weighting.
It is thus possible to simulate the SM events once (which represents the largest number of events in general), thereby saving computational time.

We have though just pointed out that the non-SM part of the cross-section can be negative.
So unless a consistent procedure for ``subtracting'' events is established, overcoming the problem of dealing with negative weights, the separation between SM and BSM contribution in the event generation is not possible.
One could simply generate everything every time, but the required computational effort makes this solution impractical.

\subsection{The importance of an $m_T$ cut and/or including the interference}
\label{sec:mtcut}

When considering hadron collisions, thus integrating the partonic cross-section (convoluted with PDFs) over the whole energy range, the interference can be the dominant contribution to the BSM signal, making the predicted total cross-section smaller than in the SM.
The pure BSM term alone cannot properly represent the new physics contribution to the total cross-section simply because the interference term contains a SM factor, which becomes large at lower energy.

In order to isolate new physics, one should cut out the low-energy region where the SM dominates (and where subsequently the interference term becomes large compared to the pure BSM contribution).
In the case of a $W^\prime$-boson search in charged lepton plus neutrino final states, this means implementing a minimum $m_T$ requirement.
This is indeed a common approach in experimental search strategies.

Observed experimental limits are currently expressed in terms of the contribution of the pure BSM term alone to the total cross-section, as a means of making them independent of any kinematic cuts.
As we just have argued, the meaning of this particular quantity is not straightforward when considering a model in which the interference does not vanish.

Furthermore what is not necessarily obvious and we therefore want to stress is how extrapolating the result over the complete range is not useful from a theoretical point of view.
If one wants to compare an observed limit with a complete BSM prediction that includes destructive interference, imposing a high enough $m_T$ cut is a necessity. 
Knowing the value of the $m_T$ cut used in the data analysis, it is in principle possible to undo the extrapolation step and obtain at least an estimate of the limit on the cut cross-section by rescaling the limit on the total cross-section according to the fraction of pure BSM contribution in the search-window.
This might only be an approximation if additional corrections enter the extrapolation procedure experimentalists use.
More importantly it is only possible to recover limits corresponding to the specific values of the $m_T$ cut used in the experimental analyses.
For some models (in general for a high enough $W^\prime$-boson mass), the provided cuts might not be optimal, i.e.\ the predicted BSM cross-section including the interference would be larger for a higher $m_T$ cut.
This theoretical cross-section might even be negative for the given cuts;
one would then have to extrapolate the provided limit to a higher $m_T$ cut in order to make a comparison,
which should lead to an underestimation of the bound on the mass of a $W^\prime$-boson.
Conversely, neglecting the interference in such a case would result in a too strict mass exclusion limit.

As the optimal $m_T$ cut strongly depends on the model, we suggest that the observed limit on the BSM cross-section should be presented \emph{as a function of} a minimum $m_T$ cut.
The theoretical prediction (including the interference) could be then accordingly computed keeping the $m_T$ cut dependence.
Note that the theoretical cross-section for the complete BSM signal is maximal when choosing the $m_T$ cut such that the differential cross-section (in $m_T$) is zero at the cut, thus integrating over the positive region only.

We conclude this section with a few additional side remarks.

As argued above, it is important to implement an $m_T$ cut, whether it is to make sure the interference can reasonably be neglected or in order to guarantee the predicted cross-section to be positive.
Going back to the issue raised in section \ref{sec:generation}, if a high enough $m_T$ cut is chosen, one can ensure the positiveness of the differential cross-section thus of the weight of the generated events.
It should therefore be possible to generate separate BSM events including the interference contribution provided the kinematic range is appropriately restricted.

If one is interested in more sophisticated search techniques involving fitting to the shape of the distribution, and no peak is clearly visible,
this implies considering the intermediate energy-range where the interference is an important contribution and cannot be ignored.

Finally, because of the interference, a reduction of events will be predicted below the region where the BSM peak starts to emerge over the background.
If the deviation is significant enough, this might actually be a useful way to probe hypothetical resonances at higher energy.
It might at least have an effect on how the background should be estimated, and thus care would be required in that respect as well.

In the next section, we illustrate our considerations by considering the $W^\prime$-boson search in the Drell-Yan leptonic channel, $pp \rightarrow W,W^\prime \rightarrow l\nu_l$, within the benchmark and the 4-site models.

\section{Heavy charged gauge bosons at the LHC}
\label{sec:application}

In this section, we give quantitative examples of the destructive interference pattern discussed in Sec. \ref{sec:interference}.
We use two reference models: the benchmark and the 4-site.
We also analyse the impact of the interference terms on $W^\prime$-boson searches at the 7 TeV LHC.

\subsection{The benchmark model}

In the benchmark model inspired by Ref. \cite{Altarelli:1989ff}, the $W^\prime$-boson is considered a heavy analogue of the SM $W$-boson with the same couplings to left-handed fermions.
Thus, $W^\prime$ decay modes and branching ratios are very similar to those of the SM $W$-boson, with the only exception being the top-bottom quark channel which opens up for $W^\prime$ masses above 180 GeV.
No mixing (or interaction) with SM gauge bosons or other heavy gauge bosons such as $Z^\prime$s is assumed.
Its width (neglecting the top-quark mass) is therefore simply
\begin{equation}
 \Gamma_{W^\prime} = \frac{4}{3} \frac{m_{W^\prime}}{m_W} \Gamma_W.
\end{equation}

As discussed in the previous section, the prediction of this model does contain interference.
One could imagine a similar model in which the interference vanishes by making the $W^\prime$ right-handed:
this would require either the inclusion of light (otherwise sterile) right-handed neutrinos or extra bosons coupling purely to right-handed quarks while still coupling to left-handed leptons.
Current analyses performed without the inclusion of any interference thus formally correspond to testing such a model instead of the benchmark model with left-handed coupling.

Within this benchmark framework, CDF \cite{Aaltonen:2010jj} and D0 \cite{Abazov:2007ah} searched for a $W^\prime$-boson in the electron-neutrino final state, and extracted a 95\% confidence level (CL) exclusion limit on the $W^\prime$ mass equal to 1.12 TeV \cite{Aaltonen:2010jj}.
Recently, searches in the combined electron-neutrino and muon-neutrino final states by both CMS and ATLAS notably extended the lower limit to: $m_{W^\prime}\ge$ 2.27 TeV and 2.15 TeV, respectively (see last publications \cite{CMS-PAS-EXO-11-024,Aad:2011yg} and previous ones \cite{Khachatryan:2010fa,Chatrchyan:2011dx,Aad:2011fe}).

These analyses are based on the production of $W^\prime$-bosons and their subsequent decay into a charged lepton (electron or muon) and a neutrino, with an individual branching fraction of about 8.5\%.
As neutrinos give rise to missing transverse momentum in the detector, the selection criteria require candidate events with at least one high transverse momentum lepton.
The off-peak, high-end tail of the SM $W$-boson production and decay constitutes the irreducible background, which is the primary source of noise.
Reducible backgrounds are also considered (see Ref. \cite{CMS-PAS-EXO-11-024} and references therein for details).
We concentrate on the signal and its irreducible background, that is on the process:
\begin{equation}
pp\rightarrow W, W^\prime \rightarrow l\nu_l 
\end{equation}
with $l=e, \mu$ and $l\nu_l=l^-\bar\nu_l+l^+\nu_l$, and at 7 TeV center-of-mass energy.
For the experimental analyses, several large Monte Carlo (MC) samples are used to evaluate signal and background efficiencies.
The MC samples for both the $W^\prime$ signal and its electroweak irreducible background are produced using PYTHIA at LO \cite{Sjostrand:2006za}.
A mass dependent k-factor for the next-to-next-to-leading order (NNLO) correction is calculated and applied to the LO cross-section.
The $W^\prime$ contribution and the irreducible background are evaluated separately as events are generated using PYTHIA, in which the interference has not been implemented for this process.

Because the neutrino cannot be measured, the experimental analysis must rely on the transverse mass of the leptonic system instead of its invariant mass, making a potential $W^\prime$ peak less prominent. Wide search-windows are thus generally used, in which the interference might matter.

\begin{figure}[ht]
\centering
\subfigure[]{
\includegraphics[width=7.7cm]{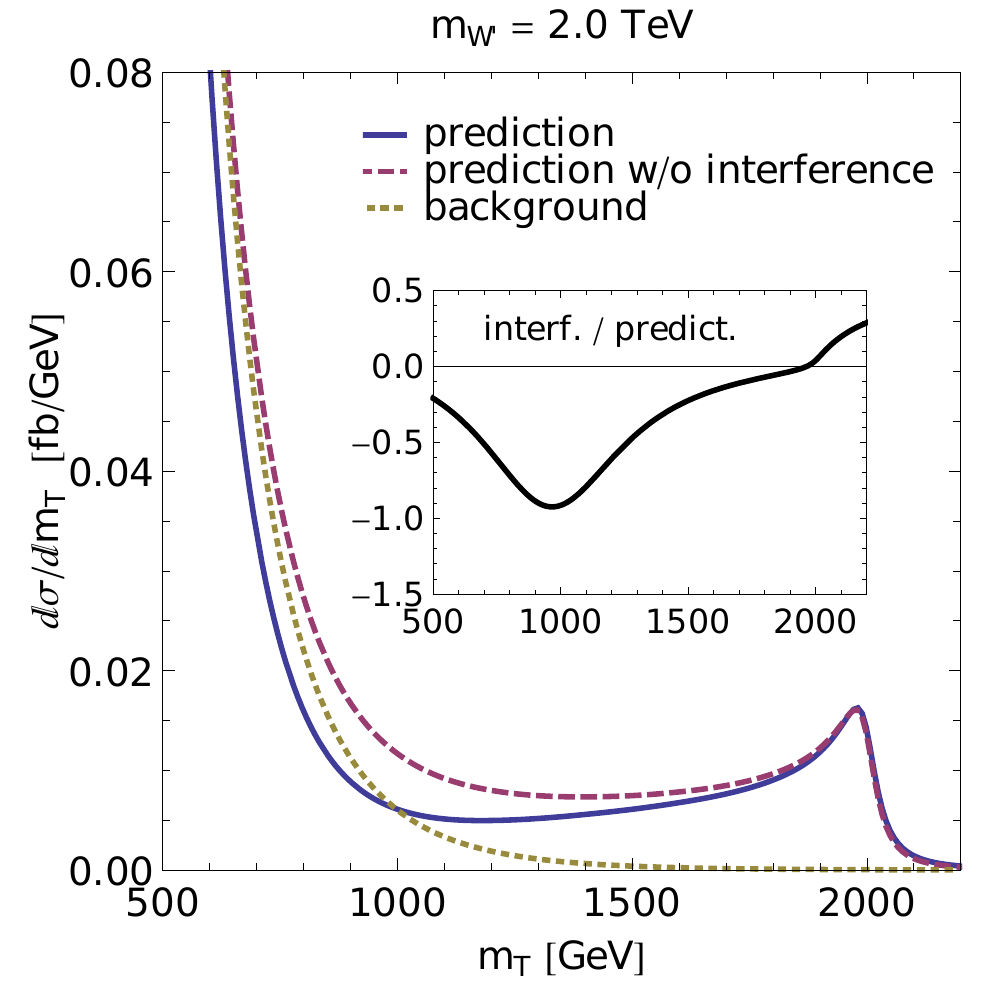}
\label{fig:totala}
}
\subfigure[]{
\includegraphics[width=7.7cm]{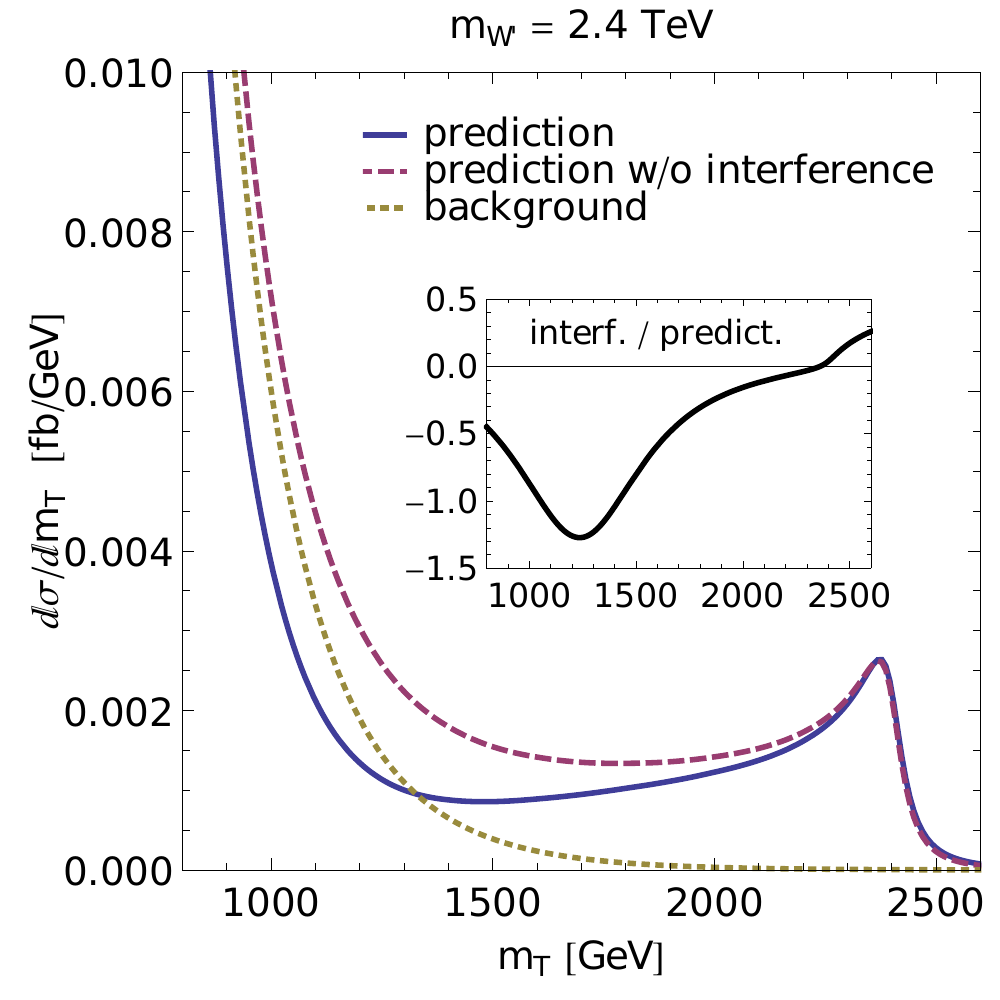}
\label{fig:totalb}
}
\caption{(Colour online)
\subref{fig:totala} Differential cross-section in the dilepton transverse mass for $m_{W^\prime}$=2000 GeV, considering 7 TeV $p p$ collisions.
The magenta dashed line shows the SM background plus the pure $W^\prime$-boson contribution as implemented in PYTHIA.
The blue solid line gives the complete theoretical distribution, including the interference term between $W$ and $W^\prime$-bosons.
The yellow dotted line represents the SM irreducible background as reference.
\subref{fig:totalb} Same for $m_{W^\prime}$=2400 GeV.
The inset plots display the interference term normalised to the complete theoretical prediction.
}
\label{fig:total}
\end{figure}

As an illustration, in Fig. \ref{fig:total} we show the differential cross-section in the electron-neutrino transverse mass, $m_T(e\nu_e)$, at LO in both electroweak and QCD interactions, for the LHC at 7 TeV (using CTEQ PDF).
We consider two representative values for the $W^\prime$-boson mass: \mbox{$m_{W^\prime}$ =} 2000, 2400 GeV.
The distribution obtained summing up SM irreducible background and pure $W^\prime$ contribution, as done in PYTHIA (magenta dashed line), is compared with the full result including the interference term (blue solid line).
The SM background is displayed as reference (yellow dotted line).
Owing to the destructive interference pattern, there is a sizeable reduction of the differential cross-section: around 1 TeV, the predicted distribution can go down to about half of what would be expected when neglecting the interference, as can be read from the inset plots.

Clearly, the inclusion of the interference term brings an important change in the shape of the $m_T$ distribution compared to the prediction obtained by summing up SM background and pure signal.
Only at around 100 GeV before the $W^\prime$-boson Jacobian peak does the interference drop down to a few percent and become negligible.
The approximation adopted by the experimental collaborations (as the interference has not been implemented in PYTHIA) has then a restricted validity domain.
Neglecting the interference outside the above-mentioned domain has three main consequences.
First of all, the choice of the optimal minimum $m_T$ cut to enhance the signal over background ratio can favour too low edges compared to the complete calculation.
Secondly, in the commonly used "counting strategy", the number of expected events collected from above the $m_T$ cut is overestimated (see Tab. \ref{tab:events} at the end of the section), which skews any estimate of the deviation from the SM prediction.
Lastly, modelling the shape of the SM irreducible background via a functional form could be biased.

\begin{figure}[ht]
\centering
\subfigure[]{
\includegraphics[width=7.7cm]{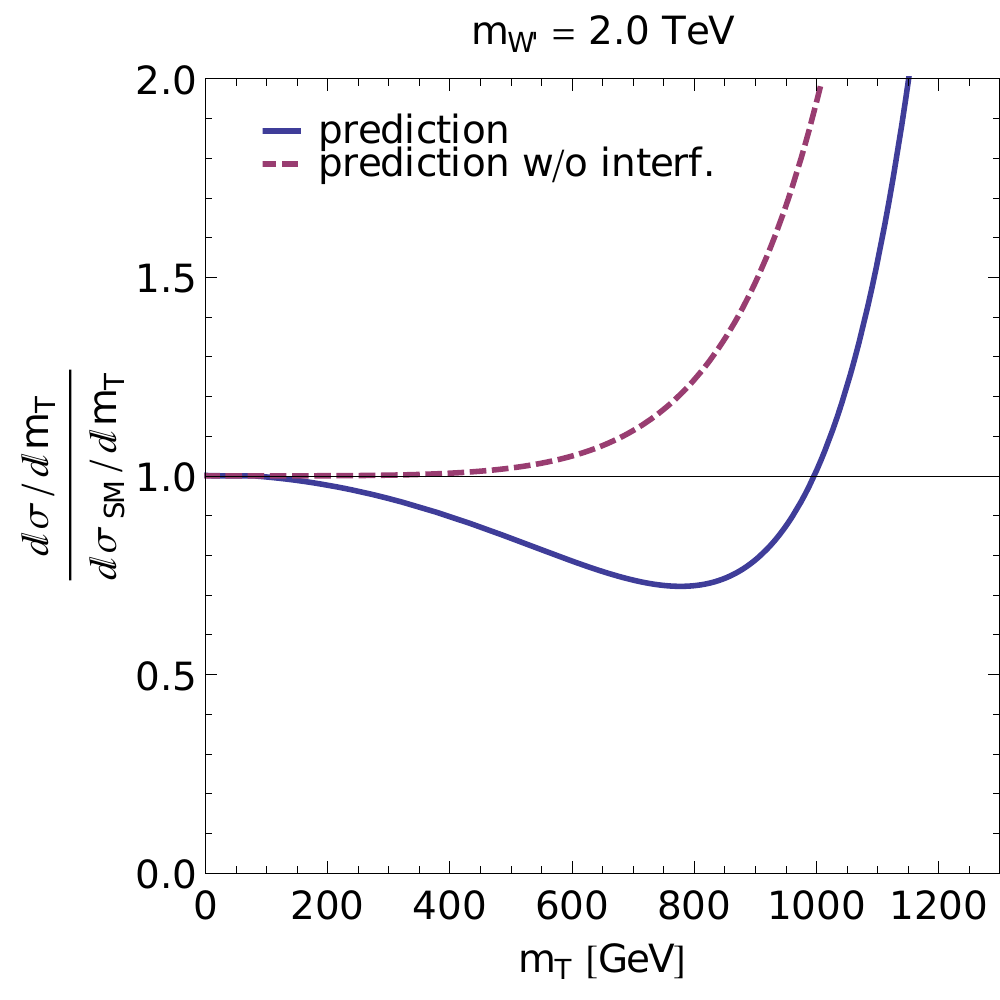}
\label{fig:ratiosa}
}
\subfigure[]{
\includegraphics[width=7.7cm]{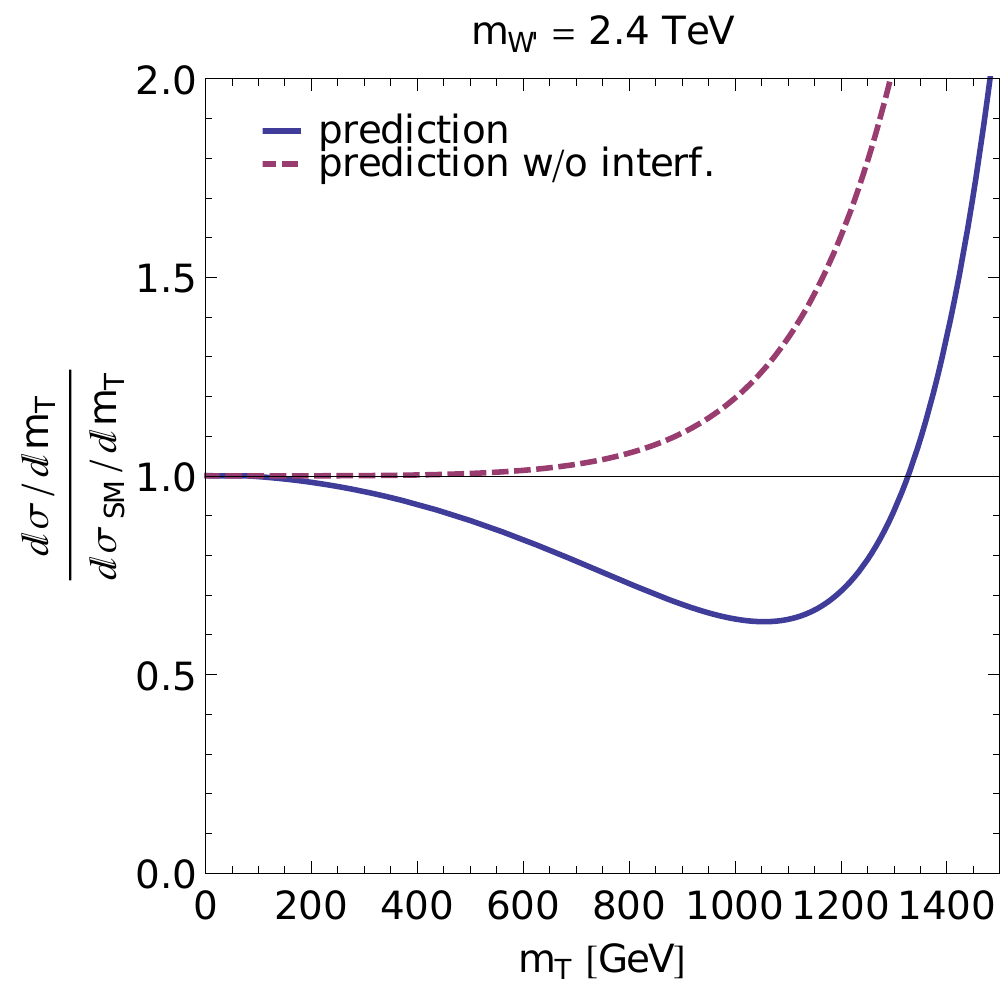}
\label{fig:ratiosb}
}
\caption{(Colour online)
\subref{fig:ratiosa} Ratio between the differential cross-section in the dilepton transverse mass for the prediction of the model and the SM irreducible background, considering 7 TeV $p p$ collisions.
The blue solid line shows the ratio obtained when including the interference.
The magenta dashed line represents the ratio without taking into account the interference.
We consider $m_{W^\prime}$=2000 GeV.
\subref{fig:ratiosb} Same for $m_{W^\prime}$=2400 GeV.
}
\label{fig:ratios}
\end{figure}

As shown in Fig. \ref{fig:ratios}, which displays the $m_T$ distribution of the model prediction normalised to the SM background,
if one neglects the interference term (magenta dashed line) the $m_T$ region which can be assumed to be BSM physics free can extend up to around 600 GeV.
Within that approximation, the BSM physics contribution is indeed apparently below the 5\% level in that range.
However, if one includes the interference (blue solid curve) the situation changes.
The $m_T$ region where the BSM physics contribution can be considered negligible shrinks down to about 300 GeV for the two representative $W^\prime$-boson masses we have chosen.
Therefore the effect of the interference should be taken into account when defining a procedure to estimate the SM background.

\begin{figure}[ht]
\centering
\subfigure[]{
\includegraphics[width=7.7cm]{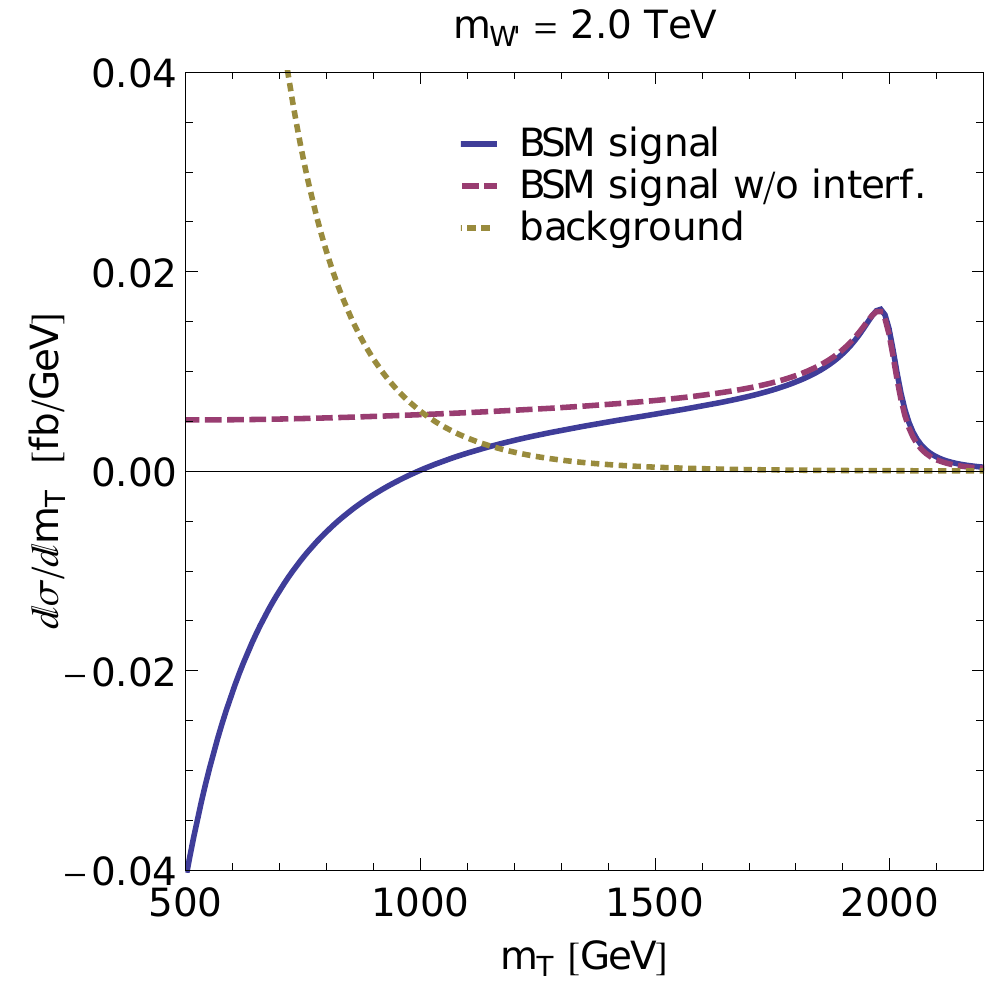}
\label{fig:signala}
}
\subfigure[]{
\includegraphics[width=7.7cm]{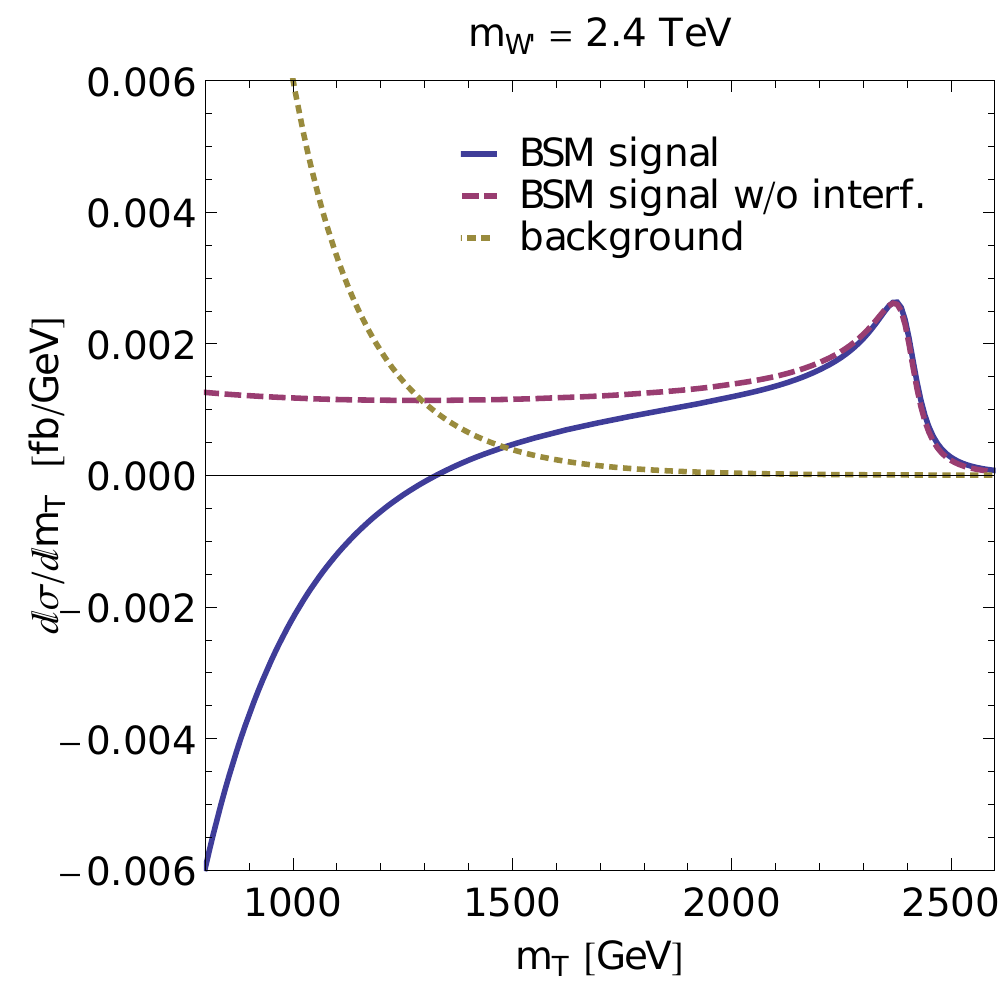}
\label{fig:signalb}
}
\caption{(Colour online)
\subref{fig:signala} Differential cross-section in the dilepton transverse mass for $m_{W^\prime}$=2000 GeV, considering 7 TeV $p p$ collisions.
The magenta dashed line shows the pure $W^\prime$-boson signal as implemented in PYTHIA.
The blue solid line gives the theoretical distribution of the BSM signal, including the interference term between $W$ and $W^\prime$-bosons.
The yellow dotted line represents the SM irreducible background as reference.
\subref{fig:signalb} Same for $m_{W^\prime}$=2400 GeV.
}
\label{fig:signal}
\end{figure}

We now concentrate more specifically on the $W^\prime$-boson signal (i.e.\ subtracting the SM irreducible background).
In Fig. \ref{fig:signal}, we plot the differential cross-section in the dilepton transverse mass for two representative values of the $W^\prime$-boson mass: $m_{W^\prime}$ = 2000, 2400 GeV.
We compare the $W^\prime$ signal prediction with and without the interference term.
As already anticipated, the complete signal becomes negative below a certain $m_T$ value.
While the approximate result is positive-definite over all the $m_T$ range.
In the LHC analyses the mass dependent k-factor for the next-to-next-to-leading order (NNLO) correction is calculated and applied to the LO ``pure'' signal cross-section.
Since the effect of the interference is significative, it would be advisable to refer instead to the computation given in Ref. \cite{Papaefstathiou:2009sr} where this term is taken into account.
We should not expect the picture to change significantly though, as the QCD corrections will affect the ``production side'' of the process, while the interference effect has to do with the propagator factor in the partonic cross-section.

\begin{figure}[ht]
\centering
\subfigure[]{
\includegraphics[width=7.7cm]{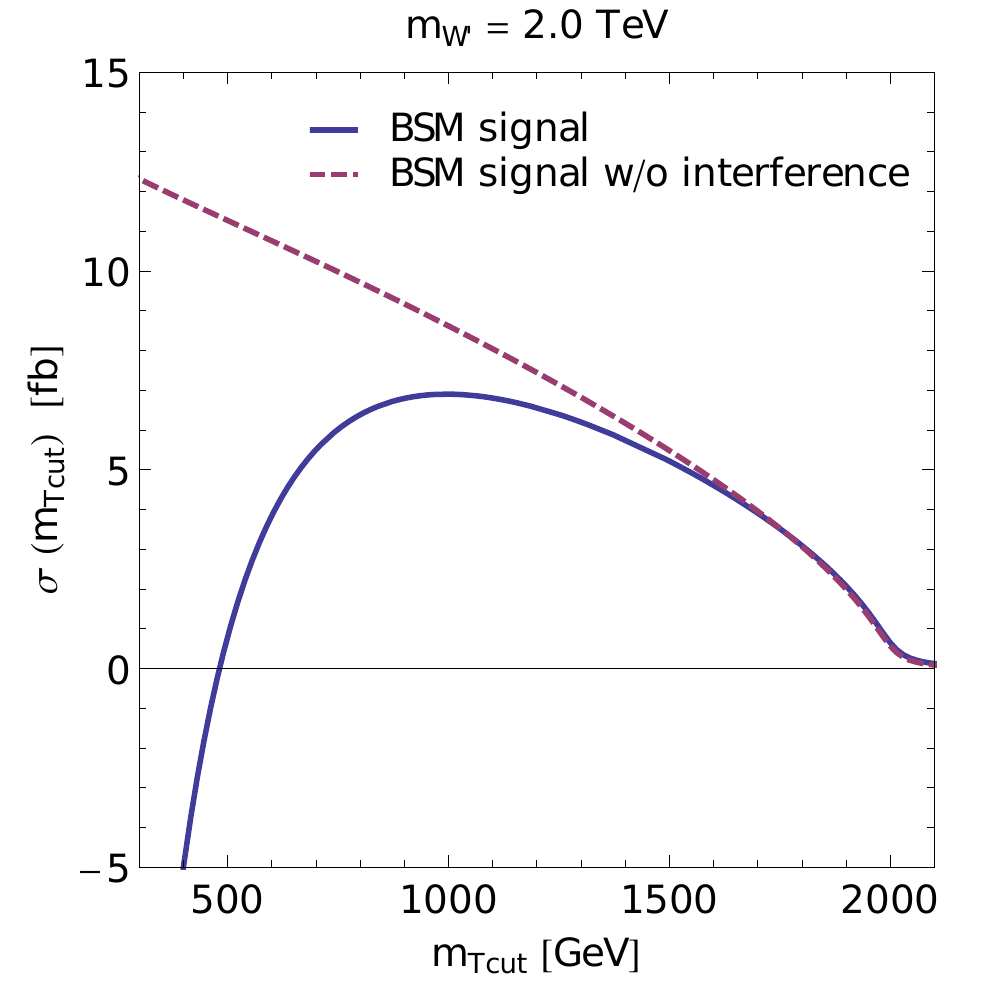}
\label{fig:signal_cumulativea}
}
\subfigure[]{
\includegraphics[width=7.7cm]{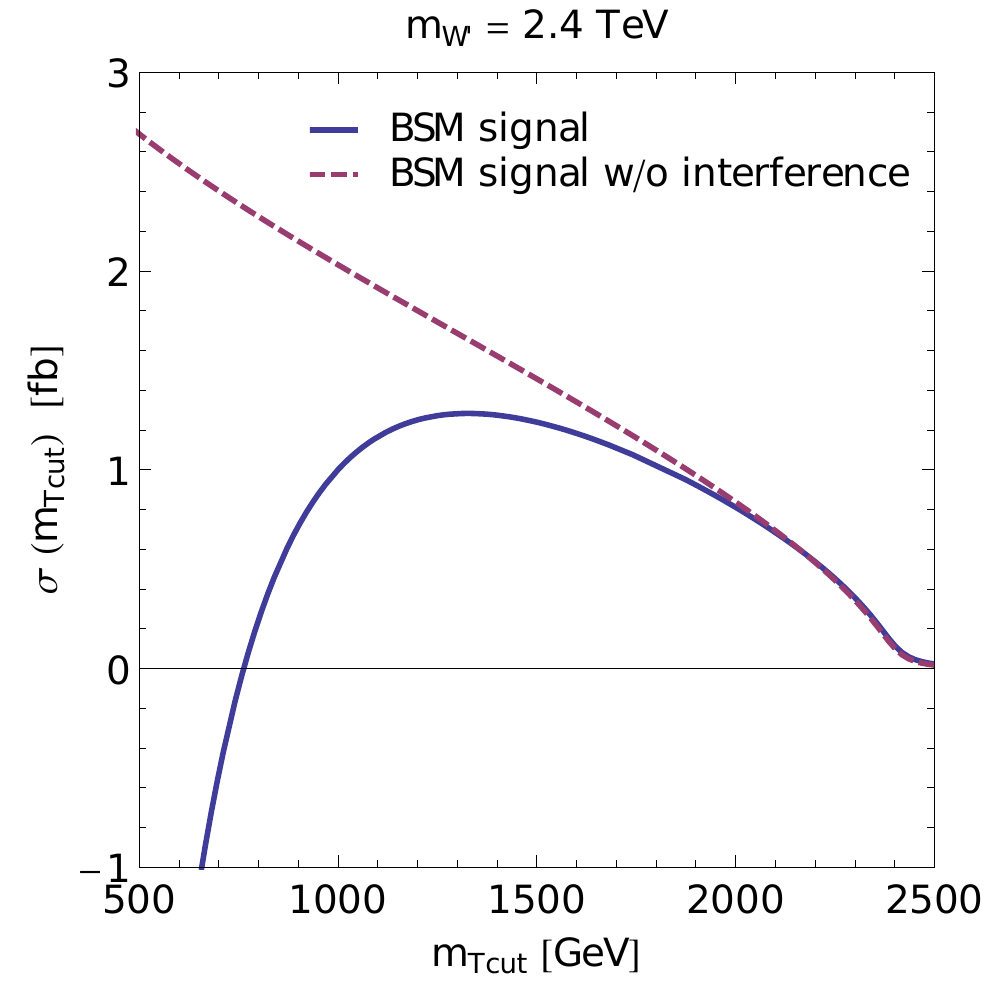}
\label{fig:signal_cumulativeb}
}
\caption{(Colour online)
\subref{fig:signal_cumulativea} $W^\prime$ signal cross-section as a function of the minimum $m_T$ cut for $m_{W^\prime}$=2000 GeV, considering 7 TeV $p p$ collisions.
The magenta dashed line shows the the pure $W^\prime$-boson contribution as implemented in PYTHIA.
The blue solid line gives the BSM signal, including the interference term between $W$ and $W^\prime$-bosons.
\subref{fig:signal_cumulativeb} Same for $m_{W^\prime}$=2400 GeV.
}
\label{fig:signal_cumulative}
\end{figure}

To quantify these effects on the integrated signal cross-section, in Fig. \ref{fig:signal_cumulative} we plot the cumulative result as a function of the lower cut on the dilepton transverse mass, ${m_T}_{\text{cut}}$.
We compare as before approximate and complete calculations for two representative values of the $W^\prime$ mass: $m_{W^\prime}$ = 2000, 2400 GeV.
As the $m_T$ cut is decreased, the divergence between the two predictions increases.
Below a critical $m_T$ cut, the new physics signal cross-section even becomes negative.
The two pieces of information one can extract from this figure are the following.

If a lower $m_T$ cut is imposed, the approximation can overestimate the signal cross-section by an amount which depends on how far from the Jacobian peak the $m_T$ cut is chosen.
This is summarised in Tab. \ref{tab:events}, where we show the overestimation of the $W^\prime$-boson signal cross-section in percent via a comparison between the approximate and complete calculations.
The values for the $m_T$ cut have been chosen according to the latest CMS analysis \cite{CMS-PAS-EXO-11-024}.
For the selected $W^\prime$-boson masses, the discrepancy can range from 4\% to 64\%.
Such an overestimation of the new physics signal leads to placing a too strict exclusion bound on the mass of a $W^\prime$.
The effect should not be too dramatic though, but big enough to be considered nevertheless.

The second piece of information is related to the existence of a critical $m_T$ cut below which the BSM signal cross-section becomes negative.
This implies, as anticipated in section \ref{sec:mtcut}, that the fully integrated signal cross-section is negative in most cases of interest (the contribution below the $W$ peak is large enough to make the total result positive only for $W^\prime$ masses below about 1.3 TeV in this model).
Values are given in the last column of Tab. \ref{tab:events}.
This illustrates the point made in section \ref{sec:mtcut} that this quantity does not reflect the prediction of the model in the search region as it can be dominated by the qualitatively different behaviour of the interference at lower energy.

\begin{table}[t]
\begin{center}
$
\begin{array}{|c||c|ccc|c||cc|}
\hline
\multirow{2}{*}{$m_{W^\prime}$}
& \multirow{2}{*}{${m_T}_{\text{cut}}$}
& \multicolumn{4}{c||}{\sigma\, ({m_T}_{\text{cut}}) ${ [fb]}$}
& \multicolumn{2}{c|}{\sigma ${ total [fb]}$}
\\ \cline{3-8}
& & ${signal}$
& ${signal}$
& ${diff.}$
& ${SM}$
& ${signal}$
& ${signal}$
\\
${[GeV]}$ & ${[GeV]}$
& ${no interf.}$
& ${with interf.}$
& ${in \%}$
& ${backgr.}$
& ${no interf.}$
& ${with interf.}$
\\ \hline
 1400 & 1000 & 67.4 & 65.0 & 3.7 & 1.1 & 131.1 & -30.1 \\
 1600 & 1100 & 31.3 & 29.7 & 5.5 & 0.6 & 60.1  & -59.3 \\
 1800 & 1100 & 16.1 & 14.6 & 10  & 0.6 & 28.5  & -63.4 \\
 2000 & 1100 & 8.0  & 6.8  & 18  & 0.6 & 14.0  & -59.0 \\
 2200 & 1100 & 3.9  & 3.0  & 32  & 0.6 & 7.1   & -52.3 \\
 2400 & 1100 & 1.9  & 1.2  & 64  & 0.6 & 3.7   & -45.6 \\ \hline
\end{array}
$
\end{center}
\caption{
From left to right, the columns indicate the $W^\prime$-boson mass value,
the minimum $m_T$ cut,
the cross-section for the $W^\prime$-boson without interference calculated from the $m_T$ lower cut on,
the cross-section for the $W^\prime$-boson with interference from the $m_T$ lower cut on,
the difference in percent between these two normalised to the latter,
the SM irreducible background from the $m_T$ lower cut on,
the total cross-section for the $W^\prime$-boson signal without and with interference.
Computed for 7 TeV $p p$ collisions.
No efficiency and acceptance factors are included.
}
\label{tab:events}
\end{table}

\subsection{The 4-site Higgsless model}

The 4-site Higgsless model represents the next-to-minimal extension of the 
3-site Higgsless model \cite{Chivukula:2006cg} that corresponds to a 
particular choice of the BESS model 
\cite{Casalbuoni:1985kq,Casalbuoni:1986vq}. They both belong to the class of 
deconstructed Higgsless theories 
\cite{ArkaniHamed:2001ca,ArkaniHamed:2001nc,Hill:2000mu,Cheng:2001vd,
Abe:2002rj,Falkowski:2002cm,Randall:2002qr,Son:2003et,deBlas:2006fz}. In their 
general formulation, these theories are based on the 
$SU(2)_L\otimes SU(2)^K\otimes U(1)_Y$ gauge symmetry, and contain K+1 
non-linear $\sigma$-model scalar fields interacting with the gauge fields, 
which trigger the spontaneous electroweak symmetry breaking. They  
constitute a viable alternative to the standard EWSB mechanism based on the 
existence of a light fundamental Higgs boson. The case K=1 corresponds to the 
minimal Higgsless model, more commonly called 3-site model. 

The 4-site Higgsless model is defined by taking K=2, and requiring the 
Left-Right (LR) symmetry in the gauge sector. More explicitly, it is a linear 
moose based on the electroweak gauge symmetry 
$SU(2)_L\otimes SU(2)_1\otimes SU(2)_2\otimes U(1)_Y$. Its theoretical 
foundations are presented in \cite{Casalbuoni:2005rs}, while some of its 
phenomenological consequences are analyzed in 
\cite{Accomando:2008jh,Accomando:2008dm,Accomando:2010ir,Accomando:2011vt}.

In the unitary gauge, the 4-site model predicts two new triplets of gauge 
bosons, which acquire mass through the same non-linear symmetry breaking 
mechanism giving mass to the SM gauge bosons. Let us denote with 
$W_{i\mu}^\pm$ and $Z_{i\mu}$ (i=1, 2) the four charged and two neutral 
heavy resonances appearing as a consequence of the gauge group extension, and 
with $W^\pm_\mu$, $Z_\mu$ and $A_\mu$ the SM gauge bosons. Owing to its gauge 
structure, the 4-site Higgsless 
model a priory contains seven free parameters: the $SU(2)_L\otimes U(1)_Y$ 
gauge couplings, $\tilde g$ and $\tilde g'$, the extra $SU(2)_{1,2}$ gauge 
couplings that, for simplicity, we assume to be equal, $g_2= g_1$, the bare 
masses of lighter ($W_1^\pm, Z_1$) and heavier ($W_2^\pm, Z_2$) gauge boson 
triplets, $m_{1,2}$, and their bare direct couplings to SM fermions, $b_{1,2}$. 
However, their number can be reduced to four, by fixing the gauge couplings 
$\tilde g,\tilde g', g_1$ in terms of the three SM input parameters 
$e, G_F, m_Z$ which denote electric charge, Fermi constant and $Z$-boson mass, 
respectively. As a result, the parameter space is completely defined by  
four independent free parameters which we choose to be $m_2$, $z$, $b_1$ and 
$b_2$, where $z=m_1/m_2$ is the ratio between the bare masses. In terms of 
these four parameters, physical masses and couplings of the extra gauge bosons 
to ordinary matter can be obtained via a complete numerical algorithm 
\cite{Accomando:2011vt}. In the following, we choose to describe the full 
parameter space via the physical observables: $m_{W2}, z, a_{W1}, a_{W2}$ which 
denote the mass of the lighter extra charged gauge boson, the ratio between 
bare masses (which, as shown in \cite{Accomando:2011vt} is a good 
approximation of the ratio between physical masses $m_{W1}/m_{W2}$), and the 
couplings of lighter and heavier extra charged gauge bosons to ordinary 
matter, respectively. 
 
In terms of the above quantities, the Lagrangian describing the 
interaction between charged gauge bosons and fermions has the following 
expression: 
\begin{equation}
\label{aw}
{\mathcal L}_{CC} = \bar{\psi} \gamma^\mu T^-  \psi \left(
a_WW_{\mu}^+ +a_{W_1} W_{1\mu }^+ +a_{W_2} W_{2\mu }^+\right) + h.c.
\end{equation}
where $\psi$ denotes SM quarks and leptons. This expression will be
used later when discussing production and decay of the two extra charged gauge 
bosons in the Drell-Yan channel. 

Any analysis is meaningless without deriving first EWPT bounds. In Ref. 
\cite{Accomando:2011vt}, we used the $\epsilon_i$ (i=1,2,3) parametrization 
\cite{Peskin:1990zt,Peskin:1991sw,Altarelli:1990zd,Altarelli:1997et} to 
extract limits on the 3-site and 4-site Higgsless models. The outcome is that 
both $\epsilon_{1,3}$ play a fundamental role in constraining the 4-site model: 
$\epsilon_3$ generates a strong correlation between the couplings of lighter 
and heavier extra charged gauge bosons to SM fermions, $a_{W1,W2}$, while  
$\epsilon_1$ limits their magnitude. Owing to the above-mentioned correlation, 
the number of free parameters can be further reduced to three. With this mild 
approximation, we can choose to describe the parameter space of the 4-site 
model in terms of the following set of physical quantities: $m_{W2}$, $a_{W2}$ 
and $z=m_1/m_2\simeq m_{W1}/m_{W2}$. At fixed $z$, the main consequences of the 
EWPT on masses and couplings of the extra heavy gauge bosons can be summarised 
as follows. Even if constrained, the $a_{W2}$ coupling can be of the same 
order of magnitude than the corresponding SM coupling. This result is common 
to all other couplings between extra gauge bosons and ordinary matter, which 
can be uniquely derived from $a_{W2}$ via our complete numerical algorithm.
This is an important property which makes a very clear distinction between 
4-site and 3-site model. The latter predicts indeed a unique gauge boson 
triplet, constrained to be (almost) fermiophobic in order to reconcile 
unitarity and EWPT bounds. Hence, oppositely to the minimal model, the 
next-to-minimal extension (or 4-site) displays the inner extra-dimensional 
nature of Higgsless theories, which are characterised by a tower of 
non-fermiophobic Kaluza-Klein resonances. The 4-site model has thus the 
potential of being detected during the early stage of the LHC experiment in 
the Drell-Yan channel. As to the spectrum, EWPT impose a lower limit on the mass
of the extra gauge bosons. The minimum mass can range between 250 and 600 
GeV, depending on the $z$-parameter value (see Ref. \cite{Accomando:2011vt} 
for computational details). The maximum allowed value for the  mass of the 
extra gauge bosons is instead fixed by the requirement of perturbative 
unitarity. As well known, one of the main motivations for Higgsless theories 
predicting an extended gauge sector, compared to the SM with no light 
elementary Higgs, is the ability to delay the perturbative unitarity violation 
up to energy scales of the order of a few TeV. Beyond that scale, new physics 
should come out. Higgsless theories must be indeed interpreted as effective 
low-energy theories. In 
\cite{Accomando:2008jh,Accomando:2008dm,Accomando:2010ir}, all vector boson 
scattering (VBS) amplitudes which are the best smoking gun for unitarity 
violations are computed, with the conclusion that the 4-site Higgsless model 
should preserve unitarity up to around 3 TeV. Thus, the allowed mass-range for the 4-site model is roughly $[250,3000]$ GeV.

A parton level analysis of the $W_{1,2}$-boson search in the Drell-Yan channel 
at the 7 TeV LHC was recently presented in Ref. \cite{Accomando:2011xi}. 
Here, we focus specifically on the effect of the interference between 
extra heavy $W_{1,2}$ and the SM $W$-bosons on the observables used in the 
experimental analysis of the final state with a charged lepton and a neutrino. 
We thus discuss how the distributions, presented in the previous section, 
appear in the 4-site model. 

\begin{figure}[ht]
\centering
\subfigure[]{
\includegraphics[width=7.7cm]{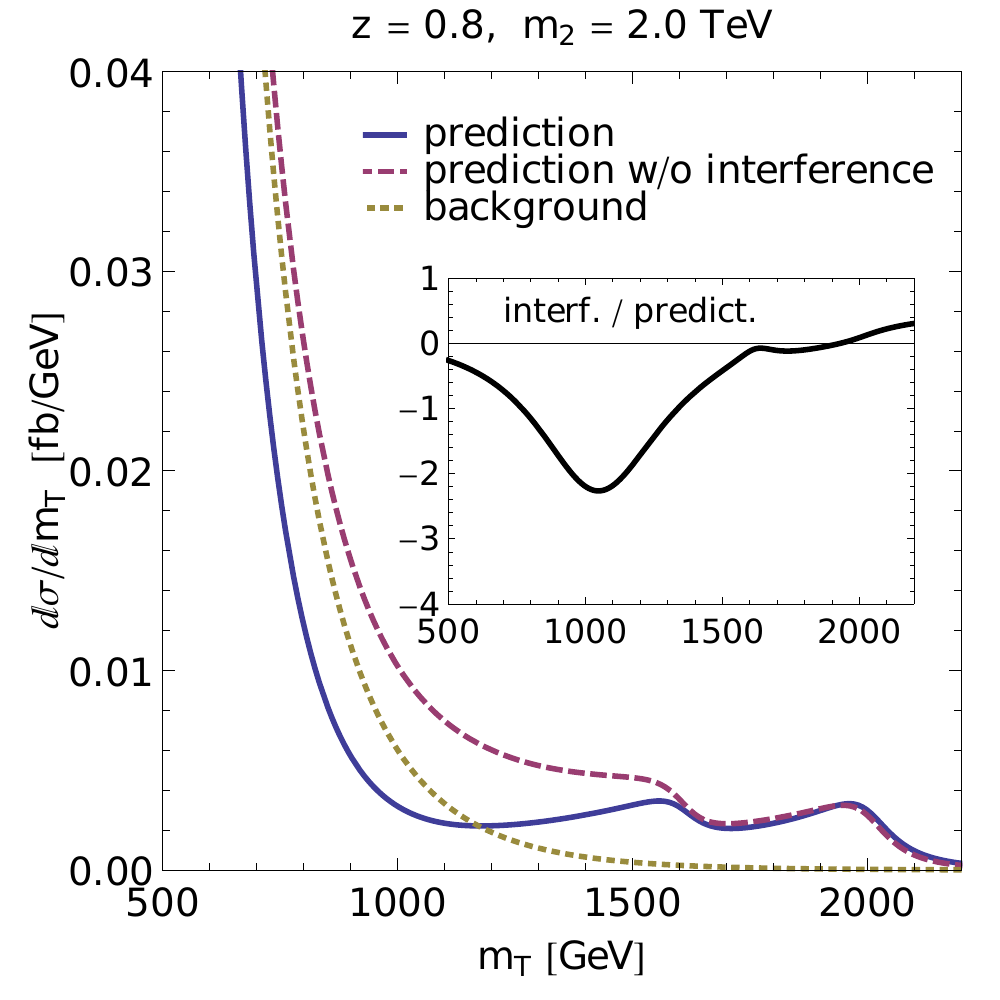}
\label{fig:total_4sitea}
}
\subfigure[]{
\includegraphics[width=7.7cm]{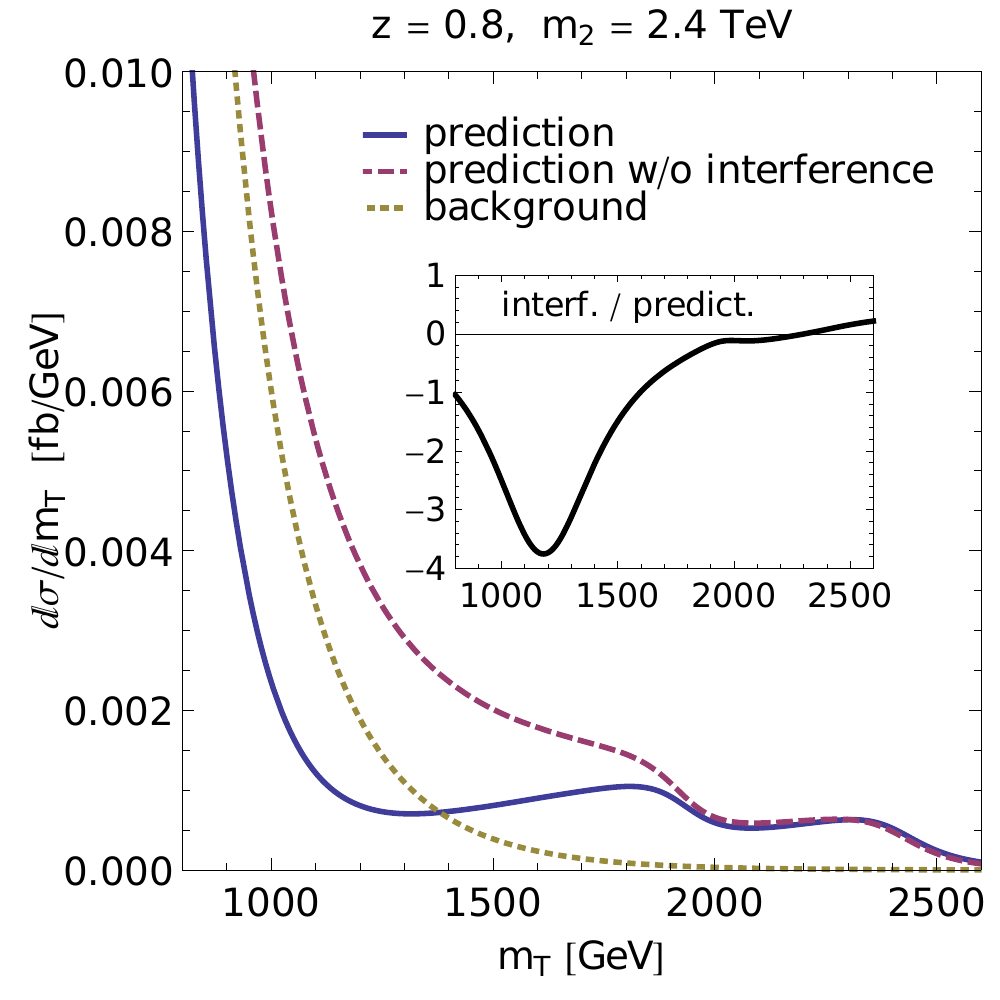}
\label{fig:total_4siteb}
}
\caption{(Colour online)
\subref{fig:total_4sitea} Differential cross-section in the dilepton transverse mass for 
$z$=0.8 and $m_{W2}$=2000 GeV, considering 7 TeV $p p$ collisions.
The magenta dashed line shows the SM background 
plus the 
pure $W_{1,2}$-boson contribution. The blue solid line gives the complete 
theoretical distribution, including the interference term between $W$ and 
$W_{1,2}$-bosons. The yellow dotted line represents the SM irreducible 
background as reference. \subref{fig:total_4siteb} Same for $m_{W2}$=2400 GeV. The inset plots
display the interference term normalised to the complete theoretical 
prediction.}
\label{fig:total_4site}
\end{figure}

In Fig. \ref{fig:total_4site}, we display the differential cross-section in 
the dilepton transverse mass, $m_T(l\nu_l)$ with $l=e,\mu$, at LO in both 
EW and QCD interactions. We choose two  
representative masses for the heavier charged gauge boson: 
$m_{W2}$ = 2000, 2400 GeV. The distribution is plotted for a fixed value of
the $z$-parameter, $z=0.8$, and for the maximal couplings between extra gauge 
bosons and SM fermions allowed by EWPT. In this setup, the mass of the 
lighter charged gauge boson is fixed to be $m_{W1}$ = 1600, 1920 GeV 
respectively. As one can see, the multi-resonance peaking structure is quite 
visible, especially when the interference is included (blue solid line).

A direct comparison with the benchmark model predictions given in the previous 
section shows that in the 4-site model the impact of the interference term on 
the $m_T(l\nu_l)$ distribution is much stronger. In the TeV region, 
the expected number of events gets indeed depleted by more than a factor of 
four with respect to the theoretical prediction without interference (magenta 
dashed line). Once again, accounting for the interference brings an important 
change in the shape of the $m_T(l\nu_l)$ distribution as compared to the 
prediction obtained by summing up SM background and pure $W_{1,2}$-boson 
contribution. In the high energy scale region used for BSM physics searches, 
only starting from the lighter resonance Jacobian peak the interference 
drops down to a few percent level, and becomes negligible as shown in the 
inset plot. This suggests that the minimum $m_T(l\nu_l)$ cut, which defines 
the search window, should be chosen around that value if one wants to work 
in the approximation where the interference is neglected. If too low, it could 
indeed bring to an overestimation of the predicted number of events within the 
adopted approximation.

\begin{figure}[ht]
\centering
\subfigure[]{
\includegraphics[width=7.7cm]{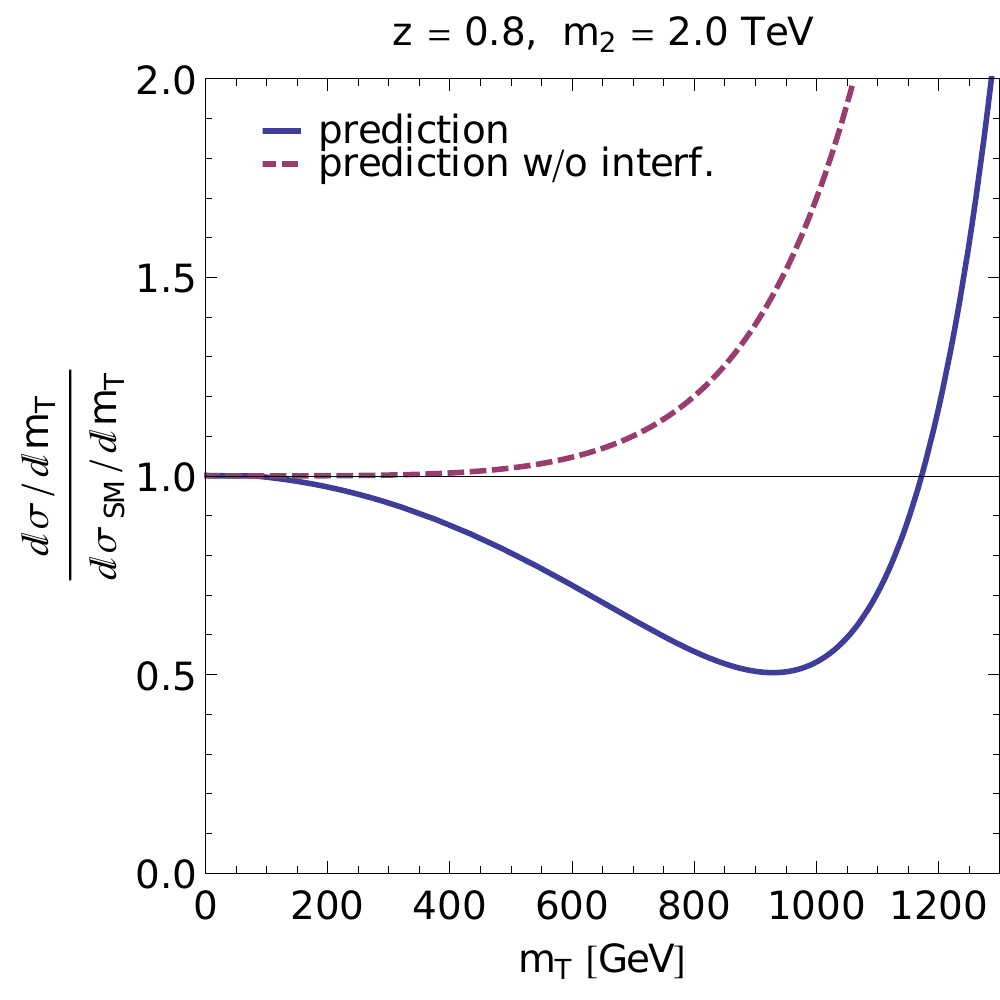}
\label{fig:ratios_4sitea}
}
\subfigure[]{
\includegraphics[width=7.7cm]{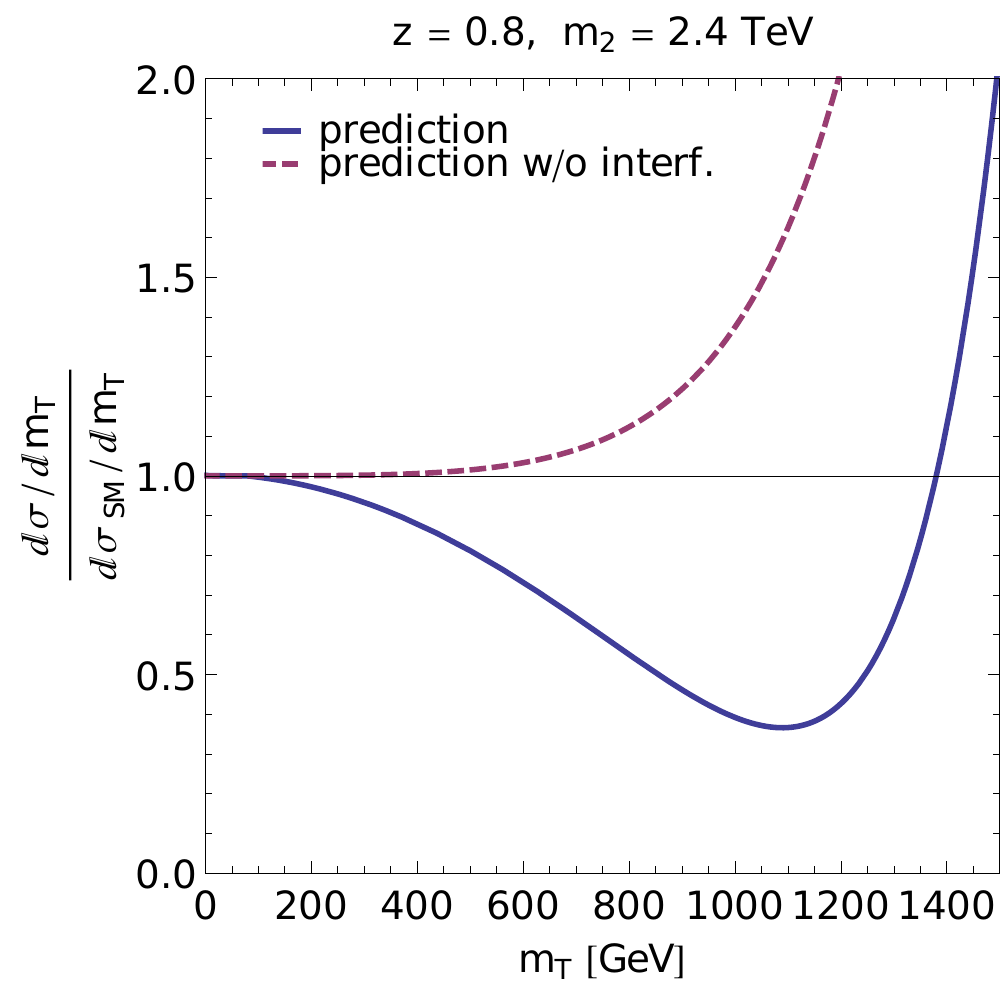}
\label{fig:ratios_4siteb}
}
\caption{(Colour online)
\subref{fig:ratios_4sitea} Ratio between the differential cross-section in the dilepton transverse mass for the prediction of the model and the SM irreducible background, considering 7 TeV $p p$ collisions.
The blue solid line shows the ratio obtained when including the interference.
The magenta dashed line represents the ratio without taking into account the interference.
We consider $z$=0.8 and $m_{W2}$= 2000 GeV.
\subref{fig:ratios_4siteb} Same for $m_{W2}$=2400 GeV.}
\label{fig:ratios_4site}
\end{figure}
Analogously, in the low $m_T(l\nu_l)$ region used for shaping the irreducible
SM background from observed data, the interference drops down to a few percent 
level only below $m_T(l\nu_l)\simeq$ 200 GeV. The BSM physics free region is 
shown in Fig. \ref{fig:ratios_4site} for $z=0.8$ and $m_{W2}$ = 2000, 2400 GeV.
As for the benchmark model presented in the previous section, here we display  
the ratio between the $m_T$ distribution for the BSM model and the
SM irreducible background. We compare the results obtained with 
(blue solid line) and without (magenta dashed line) interference terms between
$W_{1,2}$ and $W$-bosons. Also in this case, neglecting the interference leads
to an overestimation of the BSM physics free region. Within this approximation,
the $W_{1,2}$-boson contribution would indeed remain below the order of 5\% up 
to $m_T(l\nu_l)\simeq$ 600 GeV. The complete prediction suggests instead that
in order to perform a model independent analysis one should only take data in 
the restricted range $m_T(l\nu_l)\simeq$ 200 GeV to fit the functional form
of the SM background.

\begin{figure}[ht]
\centering
\subfigure[]{
\includegraphics[width=7.7cm]{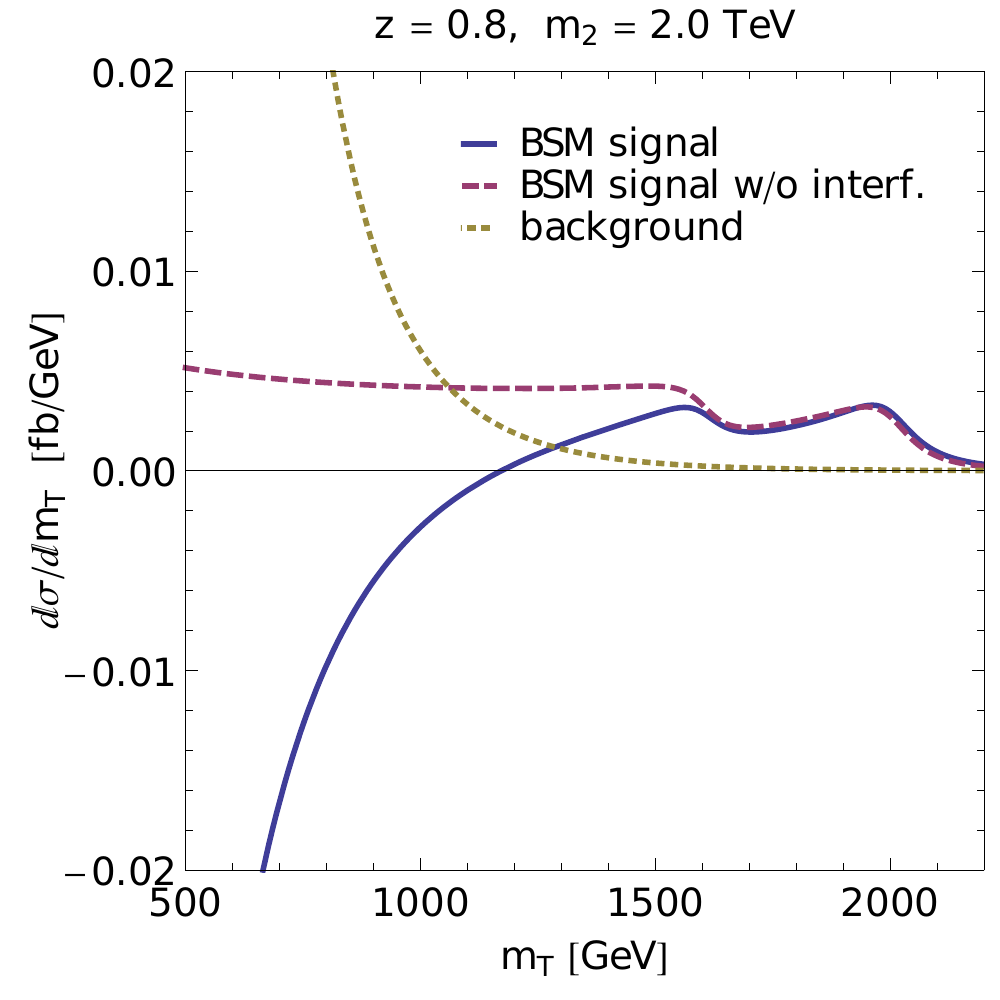}
\label{fig:signal_4sitea}
}
\subfigure[]{
\includegraphics[width=7.7cm]{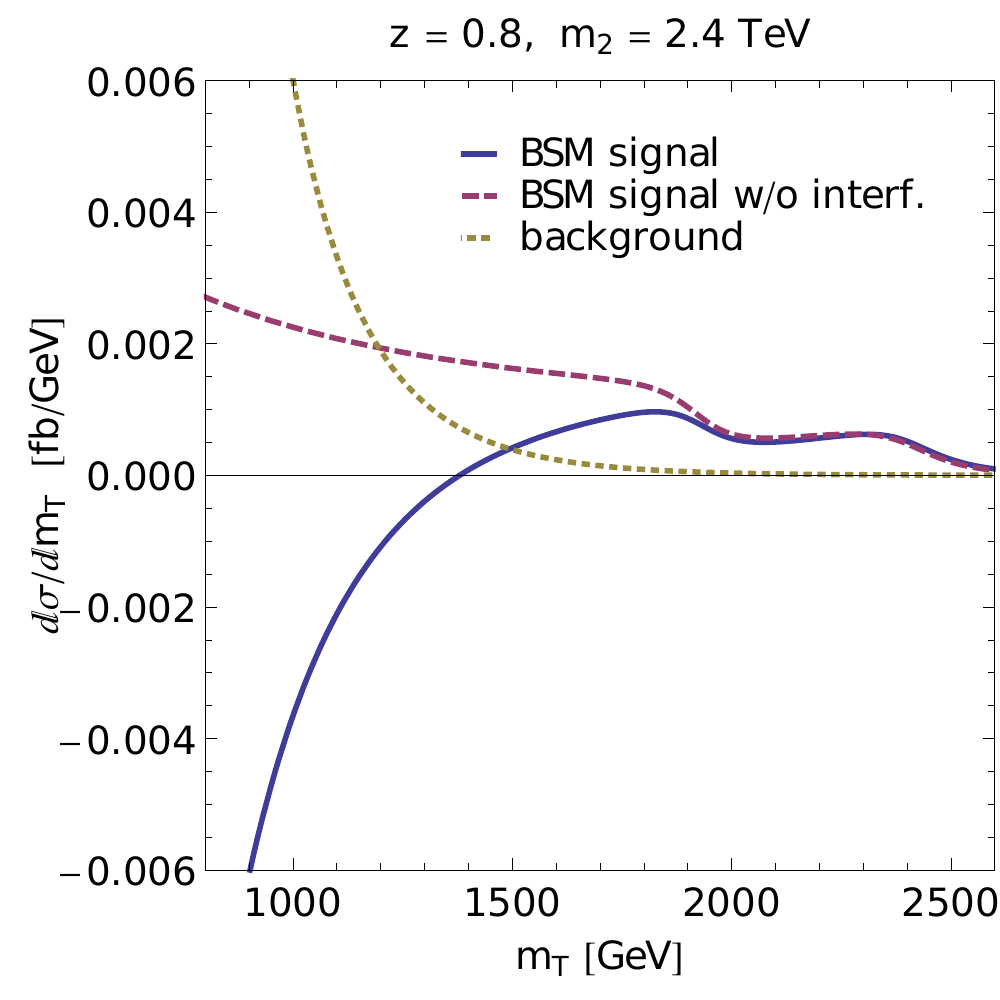}
\label{fig:signal_4siteb}
}
\caption{(Colour online)
\subref{fig:signal_4sitea} Differential cross-section in the dilepton transverse mass for $z$=0.8 and $m_{W2}$=2000 GeV, considering 7 TeV $p p$ collisions.
The magenta dashed line shows the pure $W_{1,2}$-boson contribution.
The blue solid line gives the theoretical distribution of the BSM signal, including the interference term between $W$ and $W_{1,2}$-bosons.
The yellow dotted line represents the SM irreducible background as reference.
\subref{fig:signal_4siteb} Same for $m_{W2}$=2400 GeV.}
\label{fig:signal_4site}
\end{figure}

\begin{figure}[ht]
\centering
\subfigure[]{
\includegraphics[width=7.7cm]{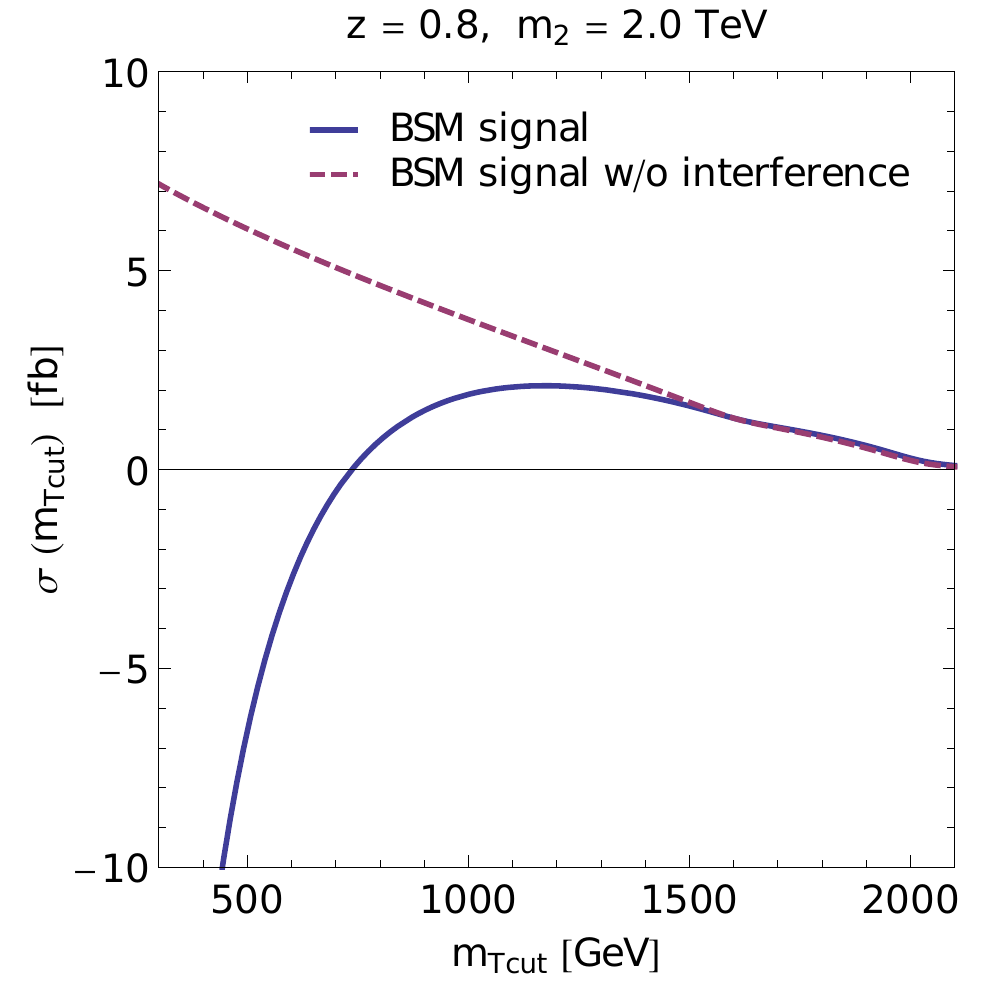}
\label{fig:signal_cumulative_4sitea}
}
\subfigure[]{
\includegraphics[width=7.7cm]{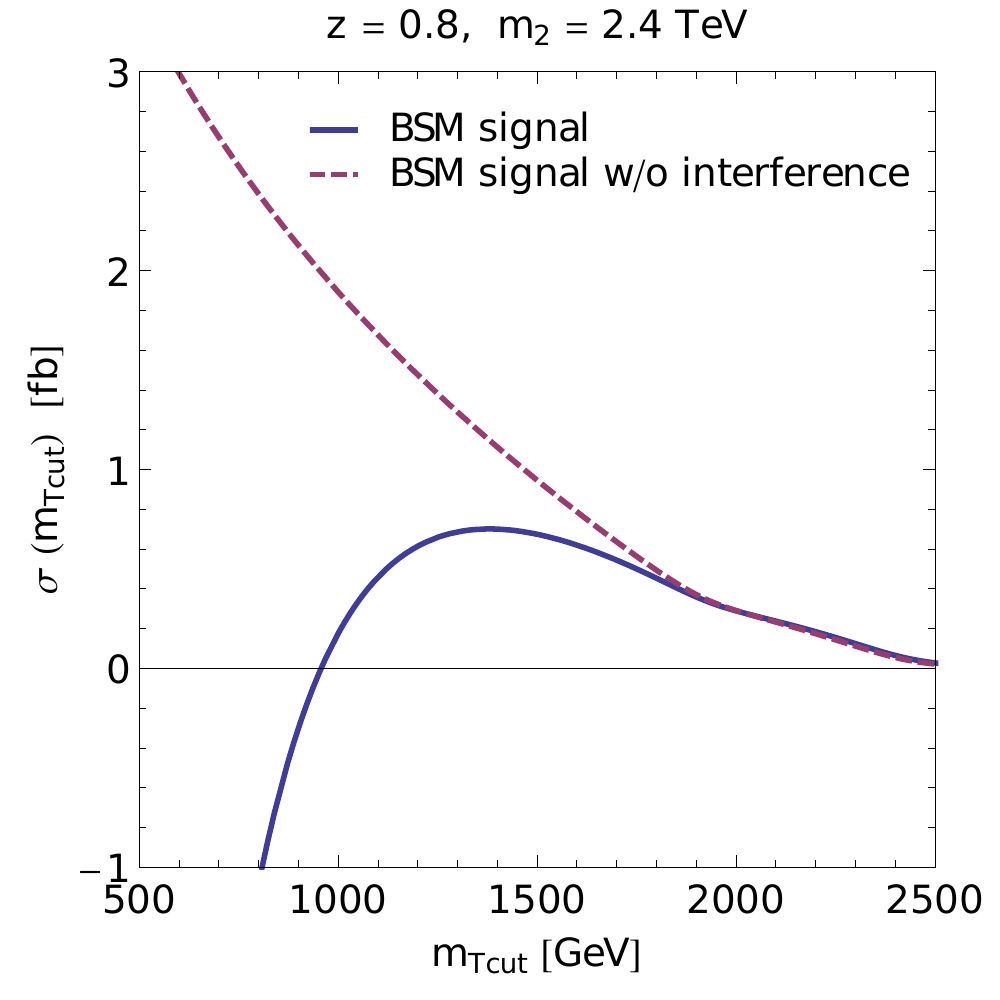}
\label{fig:signal_cumulative_4siteb}
}
\caption{(Colour online)
\subref{fig:signal_cumulative_4sitea} $W_{1,2}$ signal cross-section as a function of the minimum $m_T$ cut for $z$=0.8 and $m_{W2}$=2000 GeV, considering 7 TeV $p p$ collisions.
The magenta dashed line shows the pure $W_{1,2}$-boson contribution.
The blue solid line gives the BSM signal, including the interference term between $W$ and $W_{1,2}$-bosons.
\subref{fig:signal_cumulative_4siteb} Same for $m_{W2}$=2400 GeV.}
\label{fig:signal_cumulative_4site}
\end{figure}

We now concentrate on the interpretation of the experimental results in terms
of the 95\% CL exclusion bound on the mass of extra heavy gauge bosons.
In Fig. \ref{fig:signal_4site}, we show the $m_T$ distribution for the 
$W_{1,2}$-boson signal with (blue solid line) and without (magenta dashed line) 
interference between $W_{1,2}$ and $W$-bosons. The curves displayed support the 
result obtained previously for the benchmark model. Also within the 4-site 
model in fact the $W_{1,2}$-boson signal becomes negative below a 
certain $m_T$ value, while the approximate result is always positive-definite. 
The total cross-section for the $W_{1,2}$-boson signal can 
become negative as well. This is displayed in Fig. 
\ref{fig:signal_cumulative_4site}, where we compare the total cross-section
for the $W_{1,2}$-boson signal with (blue solid line) and without (magenta 
dashed line) interference terms between $W_{1,2}$ and $W$-bosons as a function 
of the minimum cut on the dilepton transverse mass.

The consequences of this behaviour are summarised in Tab. 
\ref{tab:events_4site}. For the minimum $m_T$ cuts imposed by recent 
experimental analyses \cite{CMS-PAS-EXO-11-024,Aad:2011yg}, the signal event 
overestimation induced by neglecting interference terms ranges between 6\% and 
266\% in the considered mass spectrum. Compared to the previously discussed 
benchmark model, the 4-site displays a much stronger impact of the interference
terms on the extraction of exclusion limits.
As shown in the last column of the table, the complete signal cross-section is negative, just as in the $W^\prime$ benchmark model, and provides another illustration of the discussion in section \ref{sec:mtcut}.
 
\begin{table}[t]
\begin{center}
$
\begin{array}{|c||c|ccc|c||cc|}
\hline
\multirow{2}{*}{$m_{2}$}
& \multirow{2}{*}{${m_T}_{\text{cut}}$}
& \multicolumn{4}{c||}{\sigma\, ({m_T}_{\text{cut}}) ${ [fb]}$}
& \multicolumn{2}{c|}{\sigma ${ total [fb]}$}
\\ \cline{3-8}
& & ${signal}$
& ${signal}$
& ${diff.}$
& ${SM}$
& ${signal}$
& ${signal}$
\\
${[GeV]}$ & ${[GeV]}$
& ${no interf.}$
& ${with interf.}$
& ${in \%}$
& ${backgr.}$
& ${no interf.}$
& ${with interf.}$
\\ \hline
 1400 & 1000 & 11.6 & 10.9 & 5.7 & 1.1 & 27.0 & -66.9 \\
 1600 & 1100 & 7.0  & 6.2  & 12  & 0.6 & 17.8 & -65.4 \\
 1800 & 1100 & 4.9  & 3.7  & 31  & 0.6 & 12.4 & -59.7 \\
 2000 & 1100 & 3.4  & 2.1  & 62  & 0.6 & 9.3  & -51.5 \\
 2200 & 1100 & 2.3  & 1.1  & 121 & 0.6 & 7.4  & -41.5 \\
 2400 & 1100 & 1.7  & 0.5  & 266 & 0.6 & 6.2  & -30.3 \\ \hline
\end{array}
$
\end{center}
\caption{From left to right, the columns indicate the $W_2$-boson mass value, 
the minimum $m_T$ cut, the cross-section for the $W_{1,2}$-boson without 
interference calculated from the $m_T$ lower cut on, the cross-section for the 
$W_{1,2}$-boson with interference from the $m_T$ lower cut on, the difference 
in percent between these two normalised to the latter, the SM irreducible 
background from the $m_T$ lower cut on, the total cross-section for the 
$W_{1,2}$-boson signal without and with interference. We assume $z$=0.8.
Computed for 7 TeV $p p$ collisions.
No efficiency and acceptance factors are included.}
\label{tab:events_4site}
\end{table}

\section{Conclusion}

In this paper, we have focussed our attention on the importance of interference  effects in searches for new physics. Our concern is that such effects are often being neglected when analysing and interpreting observed data. To make our point, we have taken as sample case the extra heavy $W^\prime$-boson search at the LHC in the charged lepton plus neutrino final state. We have considered two reference theoretical models in order to give a wider view: the so-called benchmark model inspired by Ref. \cite{Altarelli:1989ff} and used by CMS and ATLAS experimental collaborations, and the Higgsless 4-site model.

A summary examination of the matrix element squared of the $W^\prime$-boson production and decay subprocess at tree-level indicates that the
interference between SM and BSM contributions can be sizeable and negative in the intermediate energy range between the resonance peaks, in particular if the extra bosons couple universally to left-handed fermions.
To make quantitative statements about the implications of these effects, we have analysed the two above-mentioned reference models.

Our study confirms that the interference is minimal only around the Jacobian peak of the
distribution in the dilepton transverse mass $m_T(l\nu_l)$, the key observable used in experimental analyses. Away from the peak, it is substantial. It can decrease the $m_T(l\nu_l)$ differential cross-section by more than a factor of four compared to the result with no interference included. The size of this effect is model dependent and varies with the $W^\prime$-boson mass. Nevertheless, there is up to a factor of two difference already for $m_{W^\prime}\simeq $ 2.1 TeV, i.e.\ the value quoted as 95\% CL bound on the $W^\prime$-boson  mass in the latest CMS and ATLAS publications.   

Neglecting the interference has several consequences. It should affect the estimate of the optimal cut on the  $m_T(l\nu_l)$ variable to enhance the signal over background ratio, leading to an underestimation of the value of this cut.
In addition, one generally overestimates the number of predicted signal events in the search window when using this approximation.
The direct implication is that the extracted 95\% CL exclusion bound on the mass of the heavy $W^\prime$-boson in published analyses is likely to be too strict.

Another consequence concerns the  estimate of the $m_T(l\nu_l)$ range  which can be  assumed to be new physics free and thus used to derive a functional form describing the SM background via a fit to the observed data.  The derived SM background is then extrapolated to the high energy range where new physics effects are expected to appear. In the cases we have considered, the above-mentioned new physics free range shrinks by roughly a factor three when including interference terms, its upper bound going from 600 GeV down to 200 GeV.
This implies that only the range strictly around the Jacobian peak of the SM $W$-boson can formally be considered to be new physics free.
Even though points outside this range are taken into account, it can be expected that they do not contribute much to the fit, since most points lie in the lower-end of the range. 
Nevertheless, one should be aware of this effect and use appropriate care.

Lastly, the predicted signal cross-section for the production and decay of extra heavy $W^\prime$-boson(s) can be negative if the interference with SM background is included and no kinematic cut is applied.
More precisely it is negative because it is dominated by the SM contribution through the interference term, therefore is not indicative of the characteristics of the new physics.
Therefore, presenting exclusion bounds on the $W^\prime$-boson mass in terms of this observable does not have a clear direct meaning.
Currently, the interpretation of experimental results is expressed via the 95\% CL upper bound on the cross-section for the $W^\prime$-boson production and decay. This limit is cleaned up from any kinematic cuts, efficiency and acceptance factors in order to simplify its comparison with the theoretical signal cross-section and extract exclusion bounds on the $W^\prime$ mass for any possible model. This choice makes sense if the new physics contribution to the cross-section is positive-definite and independent of the SM, which is only true if the interference vanishes or if it is neglected.

Our conclusion is that the cut on $m_T(l\nu_l)$ used for the data analysis should be kept in the definition of the 95\% CL upper bound on the new physics cross-section.
Presenting limits as a function of this cut would allow for better comparisons with arbitrary models, in particular including interference effects, by providing means to choose the optimal cut case by case.

\subsection*{Note added in proof}
During the completion of this work, the CMS collaboration has released new results on the search for a new heavy gauge boson W' decaying to an electron or muon, plus a low mass neutrino \cite{Chatrchyan:2012qk}.
This study uses data corresponding to an integrated luminosity of 5.0 inverse femtobarns, collected using the CMS detector in $p p$ collisions at a centre-of-mass energy of 7 TeV at the LHC.
It determines 95\% CL mass exclusion limits for a range of W' models, also taking into account the interference between the extra W'-boson and the standard model W-boson.
Moreover, they have been following our main suggestion for presenting and interpreting data.

\begin{acknowledgments}
E.A. and D.B. acknowledge financial support from the NExT Institute and 
SEPnet. E.A. thanks the theoretical physics department of the University of 
Torino for hospitality. 
The work of S.D.C., D.D. and L.F. is partly supported by the Italian Ministero
dell'Istruzione, dell' Universit\`a
e della Ricerca Scientifica, under the COFIN program (PRIN 2008).
\end{acknowledgments}

\bibliography{references}
\bibliographystyle{utphys}

\end{document}